\documentclass[conference,compsoc]{IEEEtran}

\usepackage{tikz}
\usepackage{amsmath}
\usepackage{filecontents}
\usepackage{xcolor}
\usetikzlibrary{external, shapes.geometric, arrows}
\usepackage{colortbl}
\usepackage{threeparttable}  
\usepackage{array}
\usepackage{bbding}
\usepackage{hhline}
\usepackage{graphicx}
\usepackage{subcaption}  
\usepackage{adjustbox}
\usepackage{enumerate}
\usepackage{enumitem}
\usepackage{float}
\usepackage{booktabs}
\usepackage{caption}
\usepackage{xspace}
\usepackage{pgfplots}
\usepackage{changepage}
\usepackage{multirow}
\pgfplotsset{compat = 1.15} 
\usepackage{framed}
\usepackage{mdframed}
\usepackage[normalem]{ulem}
\usepackage{soul}
\usepackage{tabularx}
\usepackage{nicematrix}
\usepackage{threeparttable}
\usepackage{hyperref}

\usepackage{stfloats}
\usepackage{pifont}
\usepackage{multicol}

\usepackage[symbol]{footmisc}

\usepackage{makecell}
\usepackage{array, booktabs, threeparttable}

\usepackage{balance}

\newcommand{\sssec}[1]{\vspace*{0.05in}\noindent\textbf{#1}}
\newcommand{\change}[1]{\textcolor{black}{#1}}

\ifCLASSOPTIONcompsoc
  \usepackage[nocompress]{cite}
\else
  \usepackage{cite}
\fi
\ifCLASSINFOpdf
\else
\fi
\begin{document}

\title{Teaching Data Science Students to Sketch Privacy Designs through Heuristics (Extended Technical Report)$^{\star}$}

\author{\IEEEauthorblockN{Jinhe Wen$^{\diamond}$\IEEEauthorrefmark{2}, Yingxi Zhao$^{\diamond}$\IEEEauthorrefmark{2}, Wenqian Xu$^{\diamond}$\IEEEauthorrefmark{2}, Yaxing Yao\IEEEauthorrefmark{4}, Haojian Jin\IEEEauthorrefmark{2}}
\IEEEauthorblockA{\IEEEauthorrefmark{2}University of California, San Diego \IEEEauthorrefmark{4}Virginia Tech} 
}

\maketitle
\pagestyle{plain}

\begin{abstract}

Recent studies reveal that experienced data practitioners often draw sketches to facilitate communication around privacy design concepts. 
However, there is limited understanding of how we can help novice students develop such communication skills. 
This paper studies methods for lowering novice data science students' barriers to creating high-quality privacy sketches. 
We first conducted a need-finding study (N=12) to identify barriers students face when sketching privacy designs.
We then used a human-centered design approach to guide the method development, culminating in three simple, text-based heuristics. Our user studies with 24 data science students revealed that simply presenting three heuristics to the participants at the beginning of the study can enhance the coverage of privacy-related design decisions in sketches, reduce the mental effort required for creating sketches, and improve the readability of the final sketches.

\end{abstract}

\setlength{\footnotemargin}{2.5pt}

\footnotetext{\noindent\rule[0.5ex]{.285\linewidth}{0.4pt}}

\footnotetext{$^{\diamond}$Equal contributions.}

\footnotetext{$^{\star}$This report is an extended version of an IEEE S\&P 2025 paper~[\ref{ref:paper}]. We include Table~\ref{tab:ele_cover_rate_0} to further prove the improved communication efficiency of heuristic-guided privacy design sketches (Section~\ref{sec:interp-quant}). This table was skipped in the original version due to space constraints.}

\IEEEpeerreviewmaketitle

\section{Introduction}\label{sec:intro}

\begin{table*}[t]
\caption{We found that simply presenting the table below (i.e., three heuristics along with explanations and examples) to participants at the beginning of the study can enhance the coverage of privacy-related design decisions in sketches, reduce the mental effort required for creating sketches, and improve the readability of the final sketches.}\label{tab:three-principle}
\centering
\resizebox{\linewidth}{!}{
\begin{tabular}{ |m{.2\linewidth}|m{.25\linewidth}|m{.3\linewidth}| }
    \hline
    \multicolumn{1}{|c|}{\textbf{Heuristic}} & \multicolumn{1}{c|}{\textbf{Explanation}} & \multicolumn{1}{c|}{\textbf{Examples}}          \\ \hline\hline
    Device-Based Data Flow & Indicate the devices involved at each stage of the data flow to show how data moves from one point to another & Capturing photos on a camera; saving user
profiles on the server...  \\ \hline
    Stakeholder Interactions with Data Flow & Show each individual or group's interactions with the data flow, including involvement and privacy-related choices &  Manager authorizes AI modeling with user data; data scientist analyzes user's activities... \\ \hline
    Multi-Layered Representation             & Provide an overview of the privacy design, and then separate this from more detailed privacy considerations                                   & Displaying ``storing data'' in the overview, with details like ``for 5 years with encryption'' positioned in a separate layer (e.g., a different area within the sketch) \\ \hline
\end{tabular}
}
\end{table*}

\begin{figure*}[t]
            \centering
	\begin{adjustbox}{width=\linewidth,valign=t}
	\includegraphics[width=\linewidth]{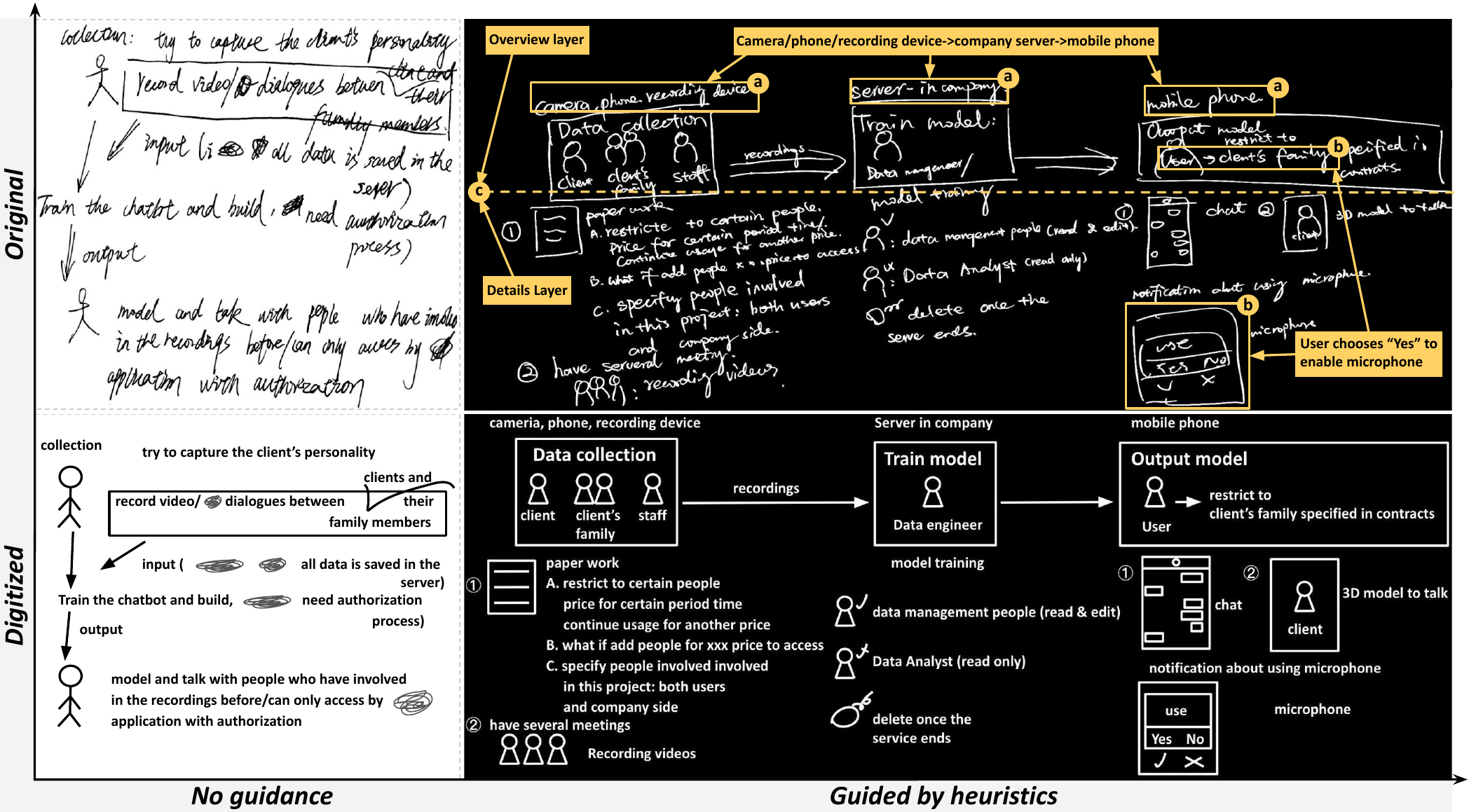}
		\end{adjustbox}

  \caption{Example sketches for the \textit{Afterlife Chatbot} scenario (see Table~\ref{tab:scenarios}), with sketches from participants exposed to heuristics (Table~\ref{tab:three-principle}) on the right and those without on the left. Heuristics-guided sketches (right) are more structured and comprehensive. The raw sketches (top) have been digitized (bottom) for improved readability.
  }\label{fig:title_fig}
\end{figure*}

Designing privacy in data practices is a collaborative process where data practitioners need to frequently communicate key privacy design concepts to others, such as how data is collected, used, shared, and how these data flows interact with users and other stakeholders~\cite{degeling2016privacy,shetty2023data}. 
A few recent studies reveal that experienced data practitioners often use sketches, either on paper or with digital tools, to facilitate communication around privacy design concepts~\cite{li2024redesigning, alhirabi2023parrot}.

Currently, there is no standardized approach to privacy sketching; instead, practitioners tend to improvise \change{in an ad-hoc manner}, combining multiple types of diagrams—such as Data Flow Diagrams, Use Case Diagrams, and Component Diagrams~\cite{alhirabi2023parrot}. This ad-hoc privacy sketching resembles user experience (UX) sketching decades ago, where only a few UX designers could sketch the UX design concepts effectively based on unstructured intuitions and artistic skills~\cite{landay1995interactive}. 
Since then, HCI researchers have developed multiple structured frameworks to lower the barriers for UX designers to sketch user experiences~\cite{buxton2010sketching}.

In this paper, we draw inspiration from UX sketching and hypothesize that structured frameworks can help lower \change{novice data science students' barriers to creating high-quality privacy sketches for given data practice scenarios.}
\change{To explore the potential framework, we conducted three studies with 54 unique participants across three U.S. universities, with no participant overlap between studies.}

We began with a need-finding study to identify the challenges novice students face in communicating privacy designs (Section~\ref{sec:needfing-method}). We recruited 12 participants, asking them to consider two data practice scenarios, sketch their privacy design solutions on paper, and explain their designs using their sketches. 
This study revealed three key challenges: (1) novice students lack the appropriate vocabulary to sketch privacy designs, (2) participants struggle with planning and organizing the sketching space, and (3) the constraints of the sketching process can limit their ability to think expansively about design solutions.

We then adopted a human-centered design approach~\cite{fitton2005rapid,harte2017human} to guide the method's development (Section~\ref{sec:teach_iter}). \change{We iteratively prototyped a teaching method,} tested it with a new group of users, gathered feedback, and refined the method iteratively. We explored three methods, ranging from a heavy-lifting approach with detailed digital diagramming to a lightweight method offering only three heuristics. Interestingly, the most lightweight method, which simply presents participants with a table of three heuristics (Table~\ref{tab:three-principle}), proved highly effective in improving the quality of novice students' sketches despite its simplicity.

We conducted a between-subjects experiment to validate the effectiveness of the heuristic-based approach (Section~\ref{sec:eval}). We recruited a total of 24 participants, with each participant creating a few sketches and interpreting the sketches from other participants. 
We found that participants who received heuristic-based instructions were able to cover 30\% more of the privacy-related decisions in their sketches than the participants without exposure to these instructions. The final sketches from the participants who received heuristic-based instruction are also more readable, with an increase of 77\% in interpretation accuracy.

Our primary contribution is a heuristic-based approach that lowers the barriers for novice students to create high-quality privacy sketches, along with insights gained from the iterative development process. 
\change{Our exploration will help researchers understand how to teach data science students to communicate privacy designs.}
To the best of our knowledge, this is the first study to systematically explore methods for teaching novice students how to sketch privacy designs.

\section{Related Work}\label{sec:rltdwork}

This project builds on ideas from four key areas: (1) user experience and software design sketches, (2) education, (3) diagrams, and (4) developer support for privacy design.

\vspace{-.195em}

\subsection{\change{User Experience and Software Design Sketches}}

Sketching has proven to be an efficient approach for exploring, communicating, and iterating on UX design ideas~\cite{greenberg2012sketching} \change{and software design concepts~\cite{cherubini2007let,branham2010let}}. 
Sketching allows designers to rapidly visualize and refine user interactions while intentionally ignoring detailed specifications in early stages \cite{landay1996silk, landay1995interactive}. This abstraction facilitates a smooth progression from initial concepts to functional design \cite{landay2001sketching}. These findings have informed the development of structured sketching practices, where designers use specific frameworks and conventions to ensure clarity and consistency in visual communication \cite{chen2003whiteboard, lewis2019sketching}. 

We \change{hypothesize} that an analogous approach can be applied to privacy design. 
Similar to how UX designers use sketches to map user journeys and identify pain points~\cite{greenberg2012sketching,landay1995interactive},
privacy practitioners can use sketching to illustrate how data is collected, processed, and shared and how the data flows interact with stakeholders.

\subsection{Education for Privacy Design}

Current privacy design education primarily focuses on privacy concepts~\cite{TrainingPBD} and design methodologies~\cite{UPSCMU, USPBerkeley,panopticon}, with most educators using case studies~\cite{UPSCMU, uwprivacyeng2020} as the main form of instruction. For instance, some classes include a privacy review procedure~\cite{SPUW, uwprivacyeng2020}, where students assess smartphone privacy notices, identify issues, and propose design improvements in brief essays. Organizations such as the National Institute of Standards and Technology (NIST)~\cite{mccallister2010identifiable} and the International Association of Privacy Professionals (IAPP)~\cite{CIPTCert} also provide privacy design education by offering standards, certifications, and online learning resources, which provide practical guidance and structure for both academic and professional learning. Additionally, educators have expanded the instruction of privacy design to a broader audience, including industry professionals and policymakers~\cite{DataPrivacyTechnology2024}. In contrast, our work aims to complement existing curricula by exploring methods to teach students how to sketch privacy designs, equipping them with privacy design communication skills.

\begin{table*}[t]
\caption{Task scenarios used in our study. \change{We initially crafted 14 privacy-related scenarios from media reports~\cite{HowRayKurzweil, ZoomCanTrack,nelsonChatrouletteReturnsHelp2021,CheatingHusbandSues2024,albrightGraphAPIKey2018,GlowPregnancyApp2020,SouthKoreanTelecom,UberUsersAre}, research papers~\cite{jin2021lean,woodruff2014would,li2024redesigning,wu2014browsemaps}, and product introductions~\cite{SetICloudMessages,ReMemoryImmortality,HereAfterAIInteractive,AppleCSAM,anomH2Oai,CompleteGuideData,CreditRiskDefinition,CoStarGroupDelivers}, then selected four scenarios based on participant familiarity, scenario complexity, and domain diversity.}}

\centering
\resizebox{\linewidth}{!}{%
\begin{tabular}{|m{.01\linewidth}|m{.28\linewidth}|m{.5\linewidth}|}
\hline
\multicolumn{1}{|c|}{\textbf{\#}} & \multicolumn{1}{c|}{\textbf{Scenario}} & \multicolumn{1}{c|}{\textbf{Description}} \\
\hline\hline
1 & Online Meeting Attention Tracking~\cite{ZoomCanTrack, li2024redesigning} & Designing an Attention Tracking feature for an online meeting app. The feature captures attendees' focus and generates attention scores, enabling hosts to monitor engagement levels during meetings. \\
\hline
2 & Financial Risk Management~\cite{anomH2Oai, CompleteGuideData, CreditRiskDefinition}  & Designing a feature for a bank’s mobile app to enhance risk management, including capabilities like anomaly detection and client credit assessments. \\
\hline
3 & Afterlife Chatbot~\cite{HowRayKurzweil,ReMemoryImmortality, HereAfterAIInteractive} & Designing an ``Afterlife Chatbot'' service that allows clients to record video biographies. After the client passes away, their family can interact with the chatbot to preserve their memory. \\
\hline
4 & Sensitive Image Detection~\cite{CoStarGroupDelivers, AppleCSAM, nelsonChatrouletteReturnsHelp2021} & Designing an automated content moderation system for an image management service to detect inappropriate or harmful content in users' uploaded pictures. \\ \hline
\end{tabular}
}
\label{tab:scenarios}
\end{table*}

\subsection{Diagrams for Privacy Design}
Practitioners often use a combination of diagrams to communicate privacy considerations in system designs \cite{li2024redesigning, alhirabi2023parrot, NistPrivacyEng2017a,jin2022exploring}. Research has shown that these diagrams help developers visualize system architecture by simplifying complex design elements, making it easier to communicate and design privacy-aware applications \cite{alhirabi2023parrot}. For example, when tasked with communicating the privacy design of an IoT application for diabetes treatment and monitoring, developers used Data Flow Diagrams (DFDs) \cite{LucidchartDFD} and Unified Modeling Language (UML) \cite{figma_uml_use_case_diagram} to break down the flow of sensitive health data into manageable components, with annotations clarifying the encryption mechanism \cite{alhirabi2023parrot}. 

While existing studies shed light on the use of diagrams for communicating privacy considerations in system design, their focus has been on professional developers. \textit{How} to teach these diagramming techniques to novice data science students remains unclear. Our work aims to address this gap by using sketches as the lens to investigate the specific challenges students face when communicating privacy design through diagrams and provide insights on targeted guidance to address these difficulties.

\subsection{Developer Support for Privacy Design} 
Prior research has explored many approaches to support experienced developers in privacy design \cite{arciniegas2017using, komanduri2011adchoices}. For instance, some studies have offered insights into how to help developers make privacy design choices by identifying the common challenges in understanding and implementing privacy considerations \cite{ayalon2017developers, leon2013matters,jin2022peekaboo}. Additionally, various tools have been developed to support developers by embedding privacy practices directly into development workflows. 
For example, Coconut, an Android Studio plugin, helps developers create privacy-friendly apps by requiring privacy annotations and assisting them in organizing privacy-related details \cite{li2018coconut}. Similarly, PARROT helps developers create privacy-aware IoT applications by providing interactive privacy annotations and guidance \cite{alhirabi2023parrot}. 

However, existing studies focus on supporting experienced developers with existing knowledge of privacy. Little attention has been given to helping novice data science students learn to communicate privacy design. Our work bridges this gap by providing targeted support to help \textbf{novices} develop the skills needed to effectively communicate privacy design.

\begin{figure*}[htb]
    \centering
    \begin{minipage}[b]{0.45\linewidth}
        \centering \includegraphics[width=\linewidth,height=5cm]{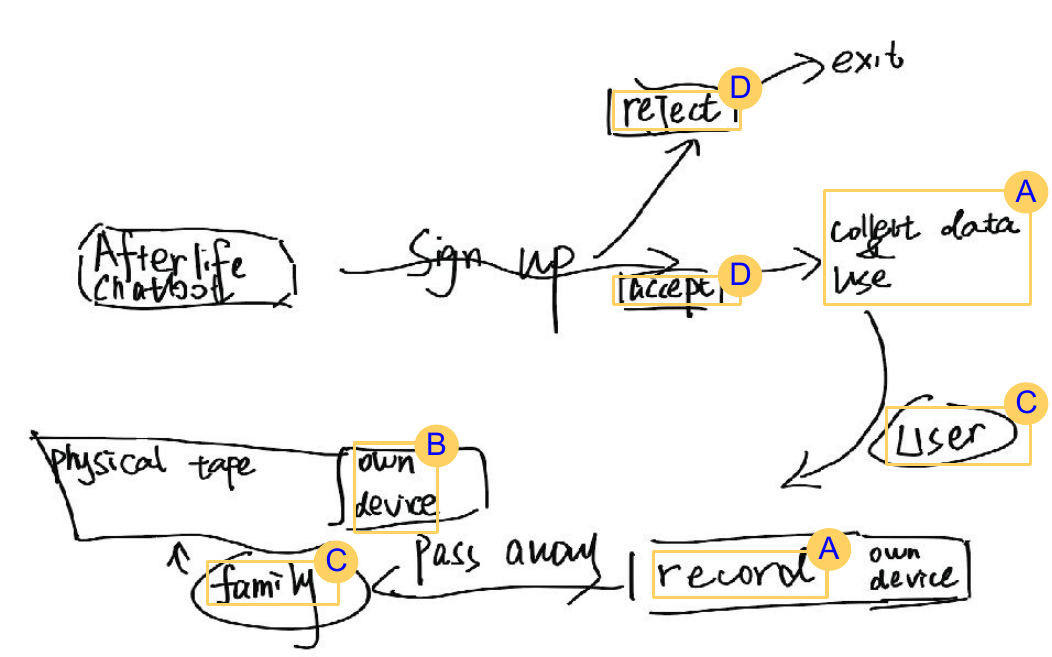}
        \subcaption{Afterlife Chatbot (by N3).}
        \label{fig:nf1}
    \end{minipage}
    \hspace{1.5cm}
    \begin{minipage}[b]{0.45\linewidth}
        \centering \includegraphics[width=\linewidth,height=5.2cm]{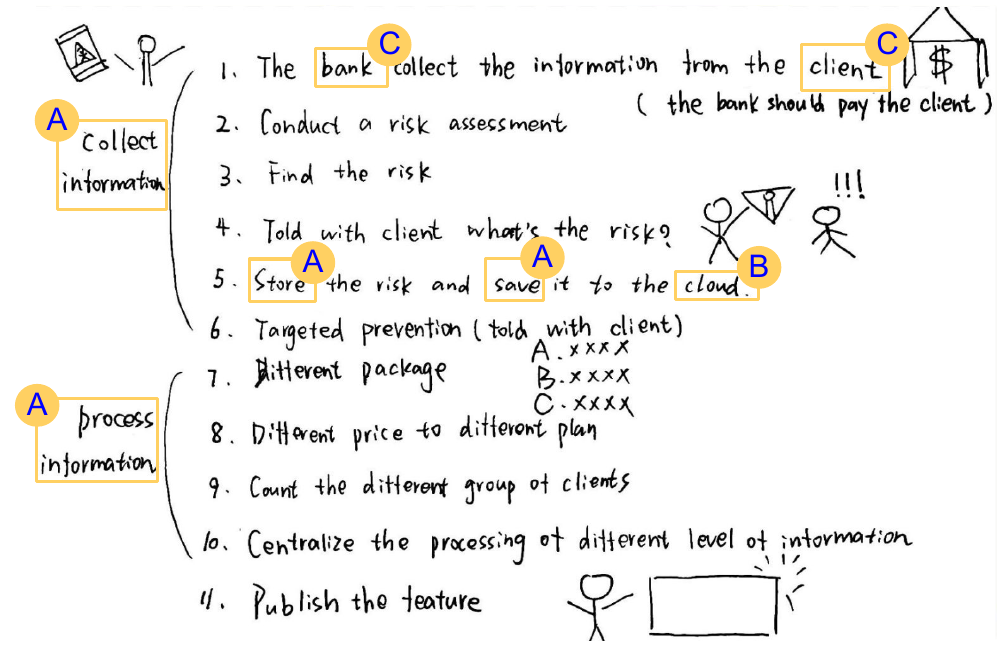}
        \subcaption{Financial Risk Management (by N6).}
        \label{fig:nf2}
    \end{minipage}
    \begin{minipage}[b]{0.42\linewidth}
        \centering \includegraphics[width=\linewidth,height=5.576cm]{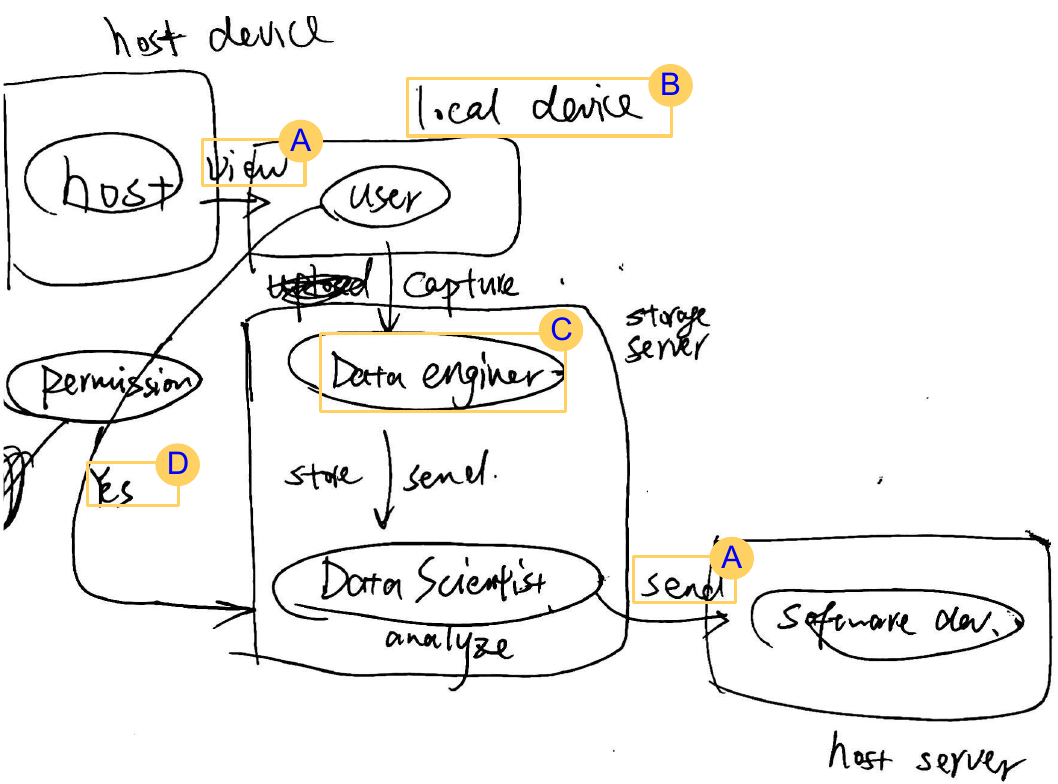}
        \subcaption{Online Meeting Attention Tracking (by N10).}
        \label{fig:nf3}
    \end{minipage}
    \hspace{1.5cm}
    \begin{minipage}[b]{0.39\linewidth}
        \centering \includegraphics[width=\linewidth,height=5.576cm]{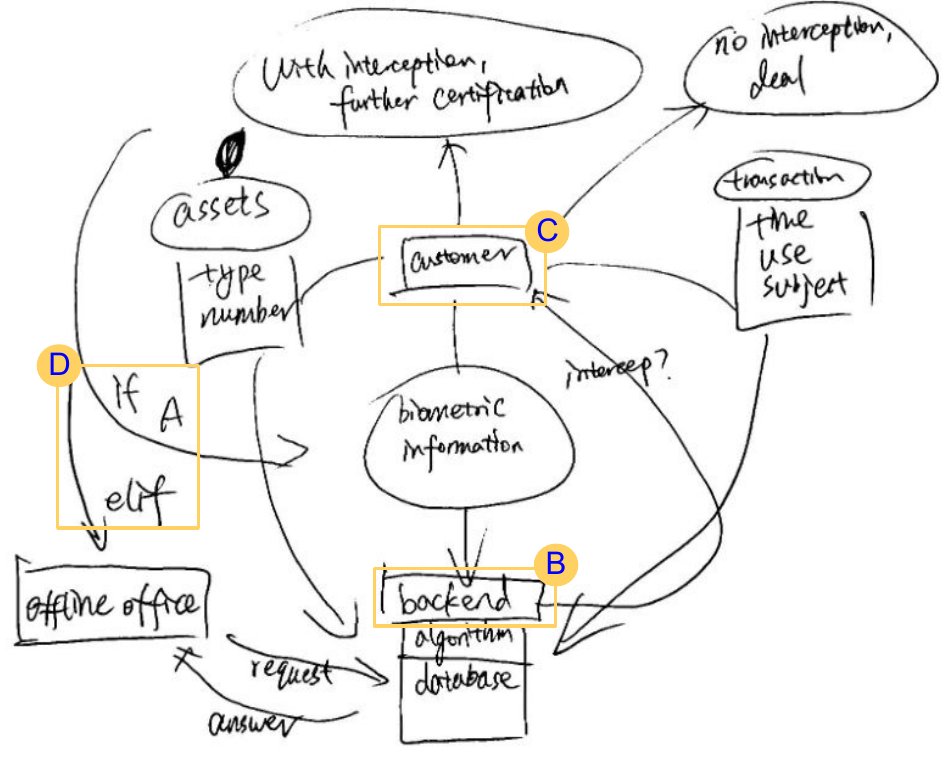}
        \subcaption{Financial Risk Management (by N11).}
        \label{fig:nf4}
    \end{minipage}
    \caption{Many sketches share some common components, which we annotate with letters: {\large{\textcircled{\small{A}}}} Data action (e.g., ``collect''); {\large{\textcircled{\small{B}}}} Device (e.g., ``server''); {\large{\textcircled{\small{C}}}} Stakeholder (e.g., ``user''); and {\large{\textcircled{\small{D}}}} Choice (e.g., ``accept''). \change{For readability, we annotate only a few examples for each component type.}
    }
    \label{fig:need-find}
\end{figure*}

\section{Why is Sketching Privacy Challenging?}\label{sec:needfing-method}

We began with a need-finding study~\cite{oulasvirta2016hci} to identify barriers for data science students in sketching privacy designs.

\sssec{Participants.}
\change{To avoid priming, we advertised the study as a ``data science experiment design study'' rather than one specifically about privacy through social media and mailing listings across three U.S. universities.}
We recruited twelve students (seven identified as female, five as male) aged 19 to 24 (Mean = 21.7 years, SD = 2.1 years). 
\change{Each participant was compensated for their time with a \$10 gift card.} 
The sample included five graduates and seven undergraduates. 
\change{We used the ACM Data Science Task Force's definition of Data Science~\cite{acmdatasci2021} to determine participants' eligibility.}

\change{We asked participants, ``Do you have any knowledge of privacy-related concepts? Please list'' in a pre-screening survey to collect information about their privacy knowledge background. Among the 12 participants, 6 answered ``No''; 3 listed relevant privacy design terms (e.g., privacy by design) they learned through research or course but lacked hands-on experience with privacy design; 3 provided responses unrelated to privacy.}

\sssec{Method.} In each study, we randomly presented students with brief descriptions of two data practice scenarios (Table~\ref{tab:scenarios}). For each scenario, we asked them to consider the data practice, sketch relevant privacy designs on paper, and explain their designs to the researchers using their sketches. To understand the fundamental barriers they face, we did not impose time limits, allowing participants to sketch freely until satisfied. 
After the sketching session, three authors reviewed the participants' sketches and asked additional questions to clarify concepts in their sketches. The authors also inquired whether they utilized any strategies or encountered any difficulties throughout their sketching process. 

On average, participants took approximately 25 minutes (SD = 5.1 minutes) to complete a sketch. 
\change{We summarized key insights after each study and identified data saturation~\cite{saunders2018saturation} by the eighth study, with no new insights emerging. We then concluded with four more participants~\cite{francis2010adequate}.}

\sssec{Findings.} We made three main observations. 
\textbf{Concepts in privacy design are abstract, and most students lack the appropriate vocabulary to sketch these concepts effectively.} 
The process of sketching privacy is the process of discovering reusable visual vocabularies. 
Interestingly, most final sketches shared common components (Figure~\ref{fig:need-find}), such as stakeholder and device, but participants were unaware of these initially. Participants often started with an arbitrary line drawing and slowly recognized the basic vocabularies through iterations, and often re-drawn the sketch multiple times near the end using a few visual components discovered in the process.
For example, when designing the Afterlife Chatbot service (the sketch without guidance in Figure~\ref{fig:title_fig}), N10 initially focused on ``client,'' ``record,'' ``video'' and ``family'' to represent an outline. However, she quickly felt lost, unsure who would manage and process these videos to build the chatbot. She also tried adding the term ``server'' to indicate where the data would be stored and used. Besides, recognizing that not everyone could view the original videos and that only family members could use the service, she included ``authorization'' to suggest access control. In the end, her sketch became a mix of arrows, lengthy text, and icons as placeholders for unspecified roles.

\textbf{Sketching privacy requires many iterations, and participants have difficulty planning the space in advance.} A simple data practice can involve many low-level privacy-related design decisions, and many of them are interdependent. 
Participants often complained that the compact positioning (e.g., Figure~\ref{fig:nf4}) makes it hard to \textit{``read through the sketch''} (N11). 
To make the problem worse, most participants need to iterate multiple rounds to have a decent design since it is hard to have a clear big picture of the privacy design before starting to sketch and think about the data practice. Participants reported that it is hard to \textit{``make targeted edits''} (N2) and complained that sketching on paper lacked the flexibility to drag-and-drop for content rearrangement. 
In the end, participants either re-started a sketch to \textit{``re-plan the layout''} (N6), squeezed updates into \textit{``any available blank space''} (N2), or even \textit{``abandoned the idea to iterate''} (N4). These challenges highlighted the need for better guidance and prompted us to explore teaching methods using digital tools (e.g., PC or iPad) to edit their already sketched content.

\textbf{The limits of the sketches mean the limits of privacy designs.} We encouraged participants to sketch while exploring the design space, but an unintended consequence emerged: they often stopped considering new design possibilities once they ran out of paper space. As a result, many sketches included surprisingly detailed components but overlooked the broader context of data practices. In the wrap-up interview, participants realized they \textit{``should include them but forgot''} (N1). Others found it physically demanding to \textit{``resketch the components''} (N2) already created in the previous steps. They expressed a desire for a \textit{``template''} (N8) to guide them on how to add details to their designs and allow for \textit{``copy-paste''} (N2) to reuse.

\begin{table*}[t]
\renewcommand{\arraystretch}{1.15}
\caption{We iteratively experimented with three methods to teach data science students sketching privacy designs. This table summarizes the design changes and experimental observations of each method.}
\resizebox{\linewidth}{!}{%
\begin{tabular}{|>{\centering\arraybackslash}m{0.13\linewidth}|>{\centering\arraybackslash}m{0.08\linewidth}|m{0.268\linewidth}|>{\centering\arraybackslash}m{0.081\linewidth}|m{0.42\linewidth}|}
\hline
\multicolumn{1}{|c|}{\textbf{Teaching Method}} & \textbf{Participants} & \multicolumn{1}{c|}{\textbf{Design Changes}}  & \multicolumn{1}{c|}{\textbf{Interface}} & \multicolumn{1}{c|}{\textbf{Observations}} \\ \hline\hline
Object-Oriented Diagramming & T1-T6 & (1) Reusable components (\texttt{DataAction} and \texttt{Stakeholder} classes); (2) Predefined attributes with preset options & Web-based application   & (1) Learning curve associated with the tool structure; (2) Constrained design options due to incomplete attributes; (3) Appreciation for the re-usable components; (4) Preference for hand-sketch tools with copy-paste support \\\hline
Vocabulary-Based Sketching& T7-T11 & (1) Visual vocabulary (e.g., ``ellipse'' for stakeholder); (2) Free text descriptions & Tablet-based sketchboard & (1) Appreciation for tablet-supported features; (2) Excessive attention to visual symbols; (3) Preference for flexible design guidelines over strict vocabulary; (4) Distraction from the overall design due to focus on details\\\hline
Heuristic-Based Sketching & T12-T18& Three heuristics: (1) Device-based data flow; (2) Stakeholder interaction with data flow; (3) Multi-layered representation & Tablet-based sketchboard & Preference for sketching with heuristics rather than vocabulary  \\\hline
\end{tabular}\label{tab:attempt-summary}
}
\end{table*}

\section{Teaching Methods Experimentation}\label{sec:teach_iter}
\sssec{Design Goal.} Our goal is to develop a method that enables data science students to create high-quality privacy sketches with minimal training. A high-quality sketch should meet two criteria: (1) it includes sufficient detail while covering the key components of the privacy design, and (2) it is clear and easy to interpret, allowing the audience to quickly understand the privacy concepts presented. 

\sssec{Experimentation Method.} 
We \change{employed an interactive prototyping method}~\cite{fitton2005rapid,harte2017human} to guide the development of the teaching method. 
\change{Interactive prototyping is a design method where users engage with evolving, often low-fidelity prototypes—such as sketches, paper interfaces, or mockups—in realistic scenarios while designers simulate system behavior, observe user interactions, offer feedback, and iteratively refine the design before the actual system is built.}
We iteratively prototyped a teaching method, tested it with a new group of users, gathered feedback, and refined the method. 
\change{Since previous research found that 80\% of usability problems are detected with four or five subjects and the most severe problems are likely to have been detected in the first few subjects~\cite{virzi1992refining}, we experimented with each method using a relatively small participant sample (5-7 for each iteration)}.

Table~\ref{tab:attempt-summary} summarizes the process and the insights we obtained through the process. 
In this section, we will describe the three methods we explored, our observations, and the trade-offs associated with each. 
\change{We recruited all participants in this experiment through the same approach as the need-finding study in Section~\ref{sec:needfing-method}. To maintain the integrity and independence of our findings, we also ensured there was no participant overlap across any of our studies, including future ones.} \change{Each participant was compensated for their time with a \$10 gift card.}

\begin{figure*}[t!]
	\centering
    \begin{minipage}[t]{0.40\linewidth}
        \centering
        \includegraphics[width=\linewidth]{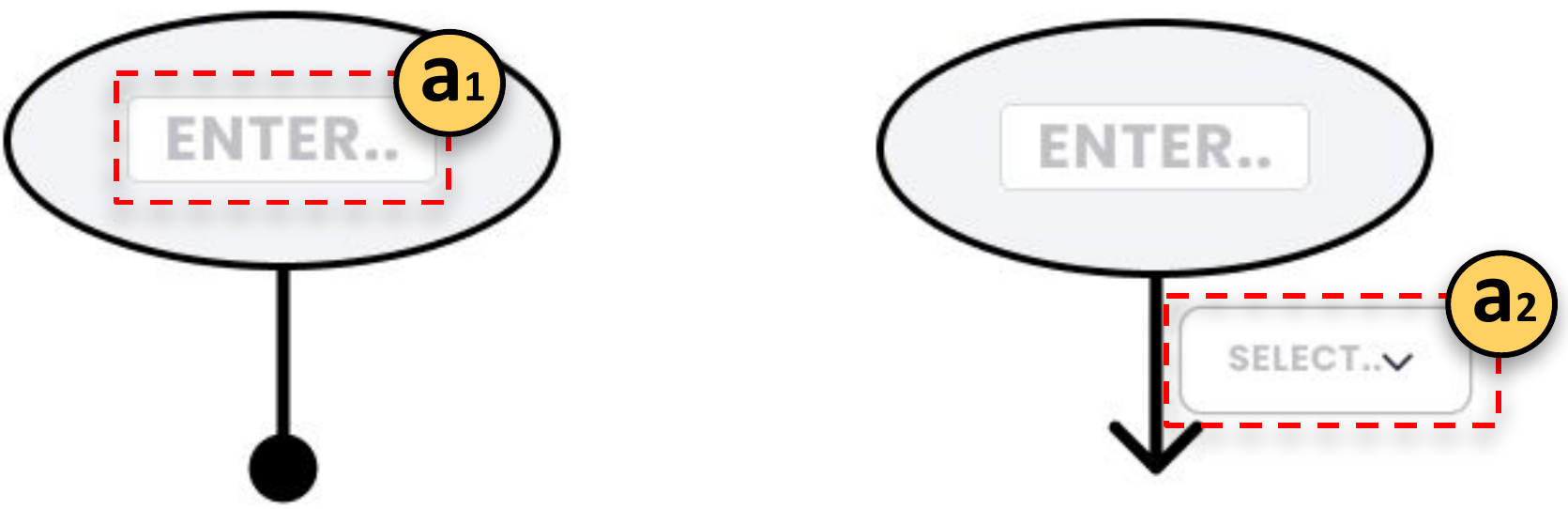}
        \subcaption{\texttt{Stakeholder} class.}\label{fig:attempt1-temp-stkhdr}
    \end{minipage}%
    \hspace{.05\linewidth}
    \begin{minipage}[t]{0.45\linewidth}
        \centering
        \includegraphics[width=\linewidth]{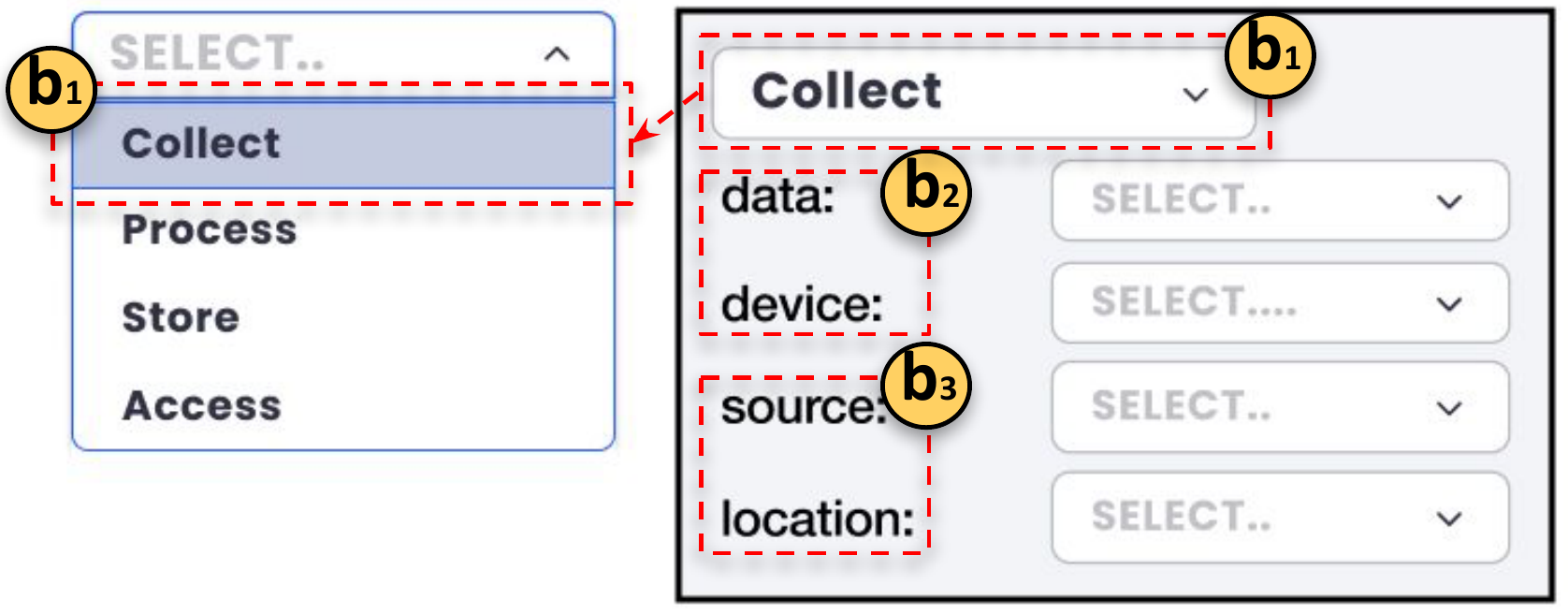}
        \subcaption{\texttt{DataAction} class.}\label{fig:attempt1-temp-action}
    \end{minipage}
    \caption{Object-Oriented Diagramming (OOD) is inspired by the Object-Oriented Programming principle in software engineering. Each basic element in OOD has methods and properties. For example, the \texttt{Stakeholder} class includes a {\large{\textcircled{\small{a$_1$}}}} name attribute for free input. Its \texttt{Involve} method (shown on the left) represents the stakeholder's involvement, while the \texttt{Decide} method (right) includes a {\large{\textcircled{\small{a$_2$}}}} binary parameter, \textit{choice}, to indicate whether a stakeholder enables or disables a \texttt{DataAction}. The \texttt{DataAction} class first requires specifying a {\large{\textcircled{\small{b$_1$}}}} 
 specific subclass. After this, students can select from preset options for both {\large{\textcircled{\small{b$_2$}}}} inherited and {\large{\textcircled{\small{b$_3$}}}} subclass-specific attributes.}\label{fig:attempt1-temp}
\end{figure*}

\begin{figure*}[ht]
    \centering
    \begin{adjustbox}{width=\linewidth,valign=t}
       \includegraphics[width=\linewidth]{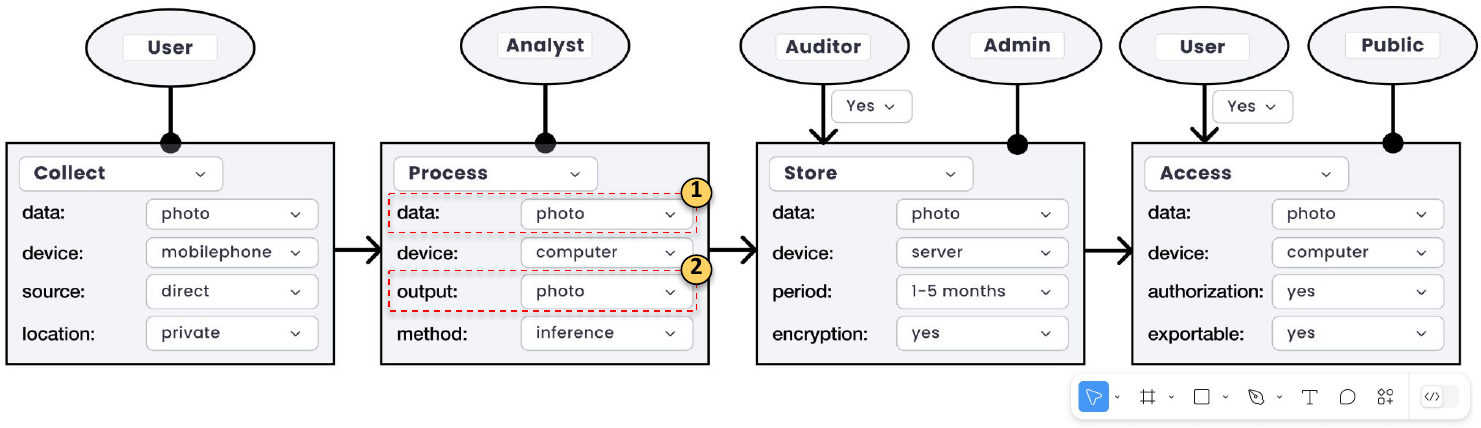}
    \end{adjustbox}
        \caption{When conducting Object-Oriented Diagramming (on Figma), participants found the reusable components helpful but experienced significant cognitive load due to the tool's learning curve and limited design options. For instance, in participant T2's design of the \textit{Sensitive Image Detection} scenario, she wanted to specify {\large{\textcircled{\small{1}}}} input and {\large{\textcircled{\small{2}}}} output data as ``raw photo'' and ``labeled photo'' for the \texttt{Process} object, but the closest option available was ``photo,'' making it challenging to communicate the design to others.}
        \label{fig:attempt1-case}
\end{figure*}

\subsection{Object-Oriented Diagramming}\label{sec:ood}

Our first attempt, Object-Oriented Diagramming (OOD), is inspired by the Object-Oriented Programming (OOP) principle in software engineering~\cite{abadi2012theory,wirfs1989object}. 

\subsubsection{Design} 
OOD has two key design ideas. 
First, we organized the privacy sketch based on the concept of objects, which can contain properties and methods (Figure~\ref{fig:attempt1-temp}) and support inheritance, similar to objects in OOP.
We defined two base classes, \texttt{DataAction} and \texttt{Stakeholder}, along with their attributes. 
\texttt{DataAction} has four sub-classes (\texttt{Collect}, \texttt{Store}, \texttt{Process}, and \texttt{Access}).
These subclasses share two common attributes, \textit{data} and \textit{device}, which indicate the direction of data flow (e.g., a photo moving from a camera to a server). Each subclass also has its specific attributes. Figure~\ref{fig:attempt1-temp-stkhdr} presents the other class \texttt{Stakeholder}, which includes a name attribute for free text input. 
It consists of two methods: \texttt{Involve} and \texttt{Decide}. \texttt{Involve} denotes a stakeholder performing a \texttt{DataAction} (e.g., an admin storing data), while \texttt{Decide} represents making a \textit{choice} (a binary parameter of the method, such as giving consent on data collection or revoking access).

Second, we offered preset attribute options to help
novice students explore the design space systematically (Figure~\ref{fig:attempt1-temp} and~\ref{fig:attempt1-case}). Instead of describing each \texttt{DataAction} in natural language, students could select predefined attribute options. For example, when specifying data \textit{source} to \texttt{Collect}, students could choose ``direct'' for newly generated data or ``secondary'' for pre-existing data.

\subsubsection{Tool Support}\label{sec:tool_support} We implemented our solution using Figma~\cite{GuideComponentsFigma}, a web application for UI design. This diagrammatic design employs three primary visual elements (Figure~\ref{fig:attempt1-temp}): \textbf{rectangle}, \textbf{ellipse}, and \textbf{edge}. Rectangles prompt students to choose from four predefined subclasses (\texttt{Collect}, \texttt{Store}, \texttt{Process}, and \texttt{Access}) with preset attribute options. Ellipses allow students to specify the stakeholder's role with custom text. Edges depict relationships: a horizontal arrow between two rectangles indicates a sequential process, while a vertical connection from an ellipse to a rectangle represents either the stakeholder's involvement (line) or a choice (arrow) that impacts the data flow.

\subsubsection{Experiment}\label{sec:attempt1-exp} 
We first tested this approach with two undergraduate and two graduate data science students. We aimed to compare the Object-Oriented Diagrams' usability and users' cognitive load with the hand-sketch method used in the pilot study (Section~\ref{sec:needfing-method}). 
Each participant created a privacy design using each approach, and we counterbalanced the order of approaches and randomly assigned two scenarios to each participant (Table~\ref{tab:scenarios}).

\change{Because the prototype is low-fidelity—for example, it does not comprehensively enumerate all properties and methods for each object—we encouraged participants to interact with the experimenters to ask questions and clarify any uncertainties during the session.}
\change{We observed that all four students experienced significant cognitive burdens when using OOD, which may have been partly due to the limited fidelity of our prototype. To address this, we recruited two additional privacy experts through personal connections to compensate for the limited fidelity—a common practice in interactive prototyping~\cite{fitton2005rapid}.}

\begin{table*}[t]
\renewcommand{\arraystretch}{1.2}
\caption{In Vocabulary-Based Sketching, we incorporate visual components and use free text descriptions to replace the previous attribute structure. For testing, we presented participants with this worksheet, including each component's name, symbol, and examples. Additionally, we specify the source and target nodes applicable to each edge component.}
\centering
\resizebox{\linewidth}{!}{%
\begin{tabular}{| m{.24\linewidth} | >{\centering\arraybackslash}m{.06\linewidth} | >{\centering\arraybackslash}m{.18\linewidth} | m{.425\linewidth} |}
\hline
\multicolumn{1}{|c|}{\textbf{Vocabulary}} &
  \multicolumn{1}{c|}{\textbf{Symbol}} &
  \multicolumn{1}{c|}{\textbf{(Source, Target)}} &
  \multicolumn{1}{c|}{\textbf{Examples}} \\ \hline\hline
Stakeholder                  & \vspace{0.05cm}\begin{tikzpicture}\draw[thick] (0,0) ellipse (0.5cm and 0.3cm);\end{tikzpicture}  & \multicolumn{1}{c|}{---}                        & User; admin; data scientist...                  \\ \hline
Data Action (on a Device) & \vspace{0.15cm}\begin{tikzpicture}
    \draw[thick] (0, 0) rectangle (1cm, 0.5cm);
    \node at (0.5cm, -0.1) {\tiny{\texttt{<DEVICE>}}};
\end{tikzpicture}  & \multicolumn{1}{c|}{---} &
  Developer captures photos on a $\langle$camera$\rangle$; engineer saves user profiles on the $\langle$server$\rangle$; data scientist analyzes user activities on their $\langle$computer$\rangle$; admin views logs on an $\langle$internal workstation$\rangle$... \\ \hline
Involvement in the Data Flow & \vspace{0.05cm}\begin{tikzpicture}\draw[-, thick] (0,0) -- (0,-0.5);\filldraw[thick] (0,-0.5) circle (2pt);
\end{tikzpicture} & (Stakeholder, Data Action)  &  See examples above (i.e., performing the action indicates the stakeholder is involved in the data flow)                                               \\ \hline
Choice on the Data Flow & \vspace{0.05cm}\begin{tikzpicture}\draw[->, thick] (0,0) -- (0,-.5cm);\end{tikzpicture}\vspace{-0.01cm} &(Stakeholder, Data Action) &    User provides consent for data collection; manager authorizes AI modeling with user data... \\ \hline
Step Procedure between Data Actions  & \begin{tikzpicture}\draw[->, thick] (0,0) -- (.5cm,0);\end{tikzpicture}\vspace{-0.01cm}  & (Data Action, Data Action) & After collecting the data, proceed to save it... \\ \hline
\end{tabular}}\label{tab:attempt2-worksheet}
\end{table*}

\subsubsection{Observations}\label{sec:attempt1-find} All participants completed their hand sketches (Mean = 25.4 minutes, SD = 4.3 minutes), and our observations aligned with from the pilot study (Section~\ref{sec:needfing-method}). However, participants experienced a significant cognitive burden when using the diagramming approach. Only one privacy expert completed the privacy design in 23.1 minutes, whereas the remaining five participants required more time, ranging from 35 minutes to over an hour. Most participants felt their privacy designs were incomplete and eventually reached a point where they couldn’t make further progress. Four participants chose to terminate the task without satisfaction after it extended beyond an hour, explaining that they \textit{``became lost''} (T3) and were \textit{``unsure how to improve the design''} (T5), for several reasons:

Participants expressed the need for more time to become familiar with the web interface and the Object-Oriented concepts. For example, one user mentioned spending a significant amount of time \textit{``navigating each component and its attributes''} (T1) to understand how it fits into their design, which they \textit{``wouldn’t need to do with a sketch''} (T4). 

Next, participants raised concerns about the \textbf{constrained design options} provided to them. For instance, all participants found the preset options of attributes limiting, including T2, who completed her diagram (Figure~\ref{fig:attempt1-case}). \textit{``I wanted to indicate the input as `original photo' and output as `labeled photo with detection results' in the processing step. However, the only option for me was `photo,' which is hard to infer if someone was reading my design.''} Others also expressed that they prefer to \textit{``use plain text to describe''} (T1) their designs. T5 complained, \textit{``This rigid structure made me feel like I was designing within a box.''} 

On the other hand, participants appreciated the \textbf{clarity of visual shapes} and the \textbf{efficiency offered by reusable components}. Many noted that preset elements significantly streamline the sketching process, as they found such elements \textit{``impractical for on-paper sketching''} (T3). Participants also expressed a strong desire to use tablets for editing efficiency. \textit{“It would be more natural to work on an iPad - I could easily duplicate and reuse components”} (T2).

\subsection{Vocabulary-Based Sketching}\label{sec:vocab}

Building on the previous observations, we refined our teaching method in design and tool support. 

\subsubsection{Design} \textbf{A Vocabulary of Privacy Design Components.} We relaxed the hierarchy requirement of the object-oriented design and simplified the vocabulary set (Table~\ref{tab:attempt2-worksheet}). For example, we eliminated the subclasses of \texttt{DataAction} and defined a Data Action node represented by a rectangle. Specifically, as one of the common components in the pilot study (Section~\ref{sec:needfing-method}), \texttt{<DEVICE>} was designed as a free text handwriting input to specify where the data action occurs. Additionally, the two methods (Figure~\ref{fig:attempt1-temp-stkhdr}) of the \texttt{Stakeholder} class and the process between data actions were represented as three types of edges connecting nodes.

\sssec{Free Text Descriptions.} To address the concern of constrained design space, we removed the design of predefined attributes. Instead, we allowed participants to use free text to describe privacy-related details.

\subsubsection{Tool Support}\label{sec:tool_support2} 
As our last experiment suggested, rather than a web-based interface, users preferred hand-sketch tools with copy-paste functionality, where they could reuse the created components. We adapted our teaching method to a tablet-based sketch board. For example, students could use an iPad with a stylus to create their privacy designs on a sketch board app such as Notability\change{~\cite{Notability}}.

\subsubsection{Experiment}\label{sec:attempt2-exp} 
We tested this approach with three undergraduate and two graduate data science students. Due to frequent difficulties in completing the diagrams (Section~\ref{sec:attempt1-find}), we did not ask them to test the Object-Oriented Diagramming in this round. We also asked each participant to sketch only one randomly selected scenario (Table~\ref{tab:scenarios}).  To support participants' learning process, we provided a worksheet (Table~\ref{tab:attempt2-worksheet}) with vocabulary, symbols, and usage examples. We guided them through its content before they started the design tasks and encouraged them to add free-text privacy details. For example, they may specify whether encryption was applied when saving user data.

\subsubsection{Observations}\label{sec:attempt2-finding}
\begin{figure*}[t]
    \centering
    \begin{adjustbox}{width=.998\linewidth,valign=t}
       \includegraphics[width=\linewidth]{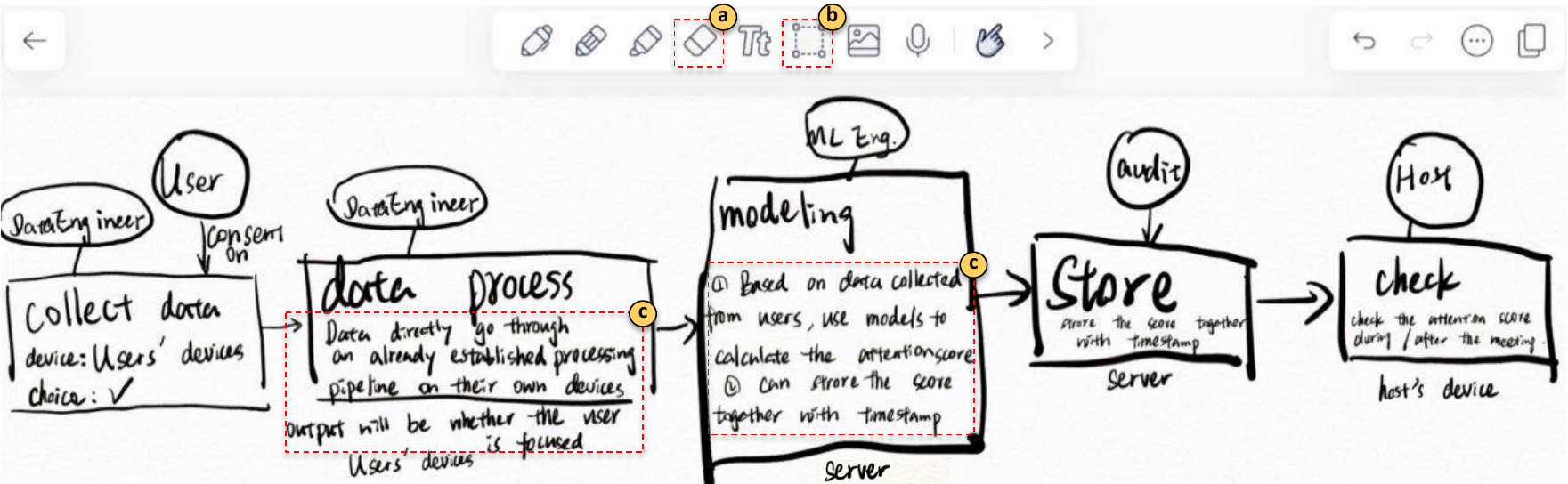}
    \end{adjustbox}
        \caption{During Vocabulary-Based Sketching, participants could use sketchboard-supported features like the {\large{\textcircled{\small{a}}}} eraser and {\large{\textcircled{\small{b}}}} selection tools (for copy-pasting and resizing) to aid their creations. However, they often became overly focused on visual symbols and were distracted by the finer details in their designs. For instance, in T11's design for the \textit{Online Meeting Attention Tracking} scenario, she included {\large{\textcircled{\small{c}}}} extensive details within each data action component, which made it challenging for her to maintain an overview of the entire design.}
        \label{fig:attempt2-case}
\end{figure*}

All participants finished their privacy design sketches (Mean = 18.6 minutes, SD = 2.7 minutes) and took advantage of the sketch board's editing features to erase, copy-paste, drag, or resize their created content (Figure~\ref{fig:attempt2-case}). However, they raised several concerns about this vocabulary-guided approach. 

First, participants reported \textbf{focusing too much on the visual components}, which they felt was \textit{``not very intuitive''} (T7). They expressed needing to \textit{``constantly stop and think about how to use each element''} (T10). Four participants wanted \textbf{flexible guidance}, with T8 saying, \textit{``Just give me a simple guideline, like including the stakeholder's engagement, so I can focus on the design without worrying about using rectangles or ellipses.''} (T8). 

Second, participants struggled to \textbf{balance }detailed privacy considerations while maintaining a design overview (Figure~\ref{fig:attempt2-case}). For instance, T10 noted, \textit{``When I was deciding how long data should be retained on the server, I lost track of the next data action I wanted to sketch.''} Three participants desired an \textit{``extra page''} to capture such details, with T9 suggesting, \textit{``An extra page for a Data Action node could include data retention periods, so I could simply edit that page if I wanted to update it from 1 month to 1 year.''}

\subsection{Heuristic-Based Sketching}\label{sec:heur}

Based on the previous findings and teaching method designs (Section~\ref{sec:needfing-method} and Table~\ref{tab:attempt-summary}), we moved toward a more lightweight solution that explores heuristics as the main approach to teach students to sketch privacy designs.

\subsubsection{Design \& Tool Support} We further simplified our method by removing constraints on vocabulary and visual representations, allowing participants to focus on the content of the design. This modification led to three heuristics (Table~\ref{tab:three-principle}) for sketching privacy designs.

\sssec{\textit{Heuristic}}{\large{\textcircled{\normalsize{1}}}}: \textbf{Device-Based Data Flow.} As one of the commonly used components in the pilot study (Section~\ref{sec:needfing-method}), the device enables students to outline their designs in a modular manner. Students can articulate data movement between devices as data actions (another common component) progress in their privacy designs. For instance, a student might sketch data flowing from a smartphone to a cloud server and then to a third-party server. This device-based communication could support further reflection on the design~\cite{frik2022users}, such as evaluating risks associated with cloud storage or potential exposure when data is shared with third-party servers.

\sssec{\textit{Heuristic}}{\large{\textcircled{\normalsize{2}}}}: \textbf{Stakeholder Interactions with Data Flow.} This heuristic encourages students to approach their design from a role-based perspective, incorporating each party's roles in the privacy design. Here, ``interaction'' could include any privacy-related behavior of stakeholders, such as their involvement and decisions about the data flow. For example, a student might sketch an admin accessing user data, which is then shared with a data scientist for further analysis. By articulating stakeholders within their design, designers and their collaborators could trace accountability~\cite{nissenbaum1996accountability}, ensuring stakeholders are held responsible for unauthorized access or data use beyond its intended purpose.

\sssec{\textit{Heuristic}}{\large{\textcircled{\normalsize{3}}}}: \textbf{Multi-Layered Representation.} Inspired by DENIM~\cite{newman2003denim}, a multi-layered approach to website design that supports quick updates and helps maintain awareness of the overall structure, this heuristic encourages students to work with both high-level and detailed perspectives in their privacy designs. By using this multi-layered approach, students can plan their privacy design before diving into specifics and refer back to the overview to maintain a clear sense of direction throughout their design process.

\subsubsection{Experiment}\label{sec:attmpt3-exp} We tested the Heuristic-Based Sketching against Vocabulary-Based Sketching with four undergraduate and three graduate data science students. Each participant sketched privacy designs using both approaches, and we counterbalanced the order of approaches and two scenarios presented to them (\textit{Afterlife Chatbot} and \textit{Financial Risk Management}, see Table~\ref{tab:scenarios}). \change{We chose these two scenarios because previous participants found these topics interesting, which enhances participants’ engagement.}

To support their learning process, we included a warm-up session before each task based on feedback from previous participants (Section~\ref{sec:attempt2-finding}). We introduced an example scenario (\textit{Online Meeting Attention Tracking}, see Table~\ref{tab:scenarios}) and guided them through the corresponding worksheet (Table~\ref{tab:three-principle} or~\ref{tab:attempt2-worksheet}). Participants then sketched the example scenario and asked any questions only during warm-up.

\subsubsection{Observations}\label{sec:attmpt3-find} 
All participants completed the warm-up within 20 minutes (Mean = 18.5 minutes, SD = 1.2 minutes) and reached satisfaction with each design within 15 minutes (Mean = 13.6 minutes, SD = 2.9 minutes). First, we observed a significant \textbf{reduction in completion time} with the vocabulary-based approach (14.3 vs. 18.6 minutes in Section~\ref{sec:attempt2-finding}), as participants reported that the warm-up session helped them become familiar with the method. Additionally, there was a clear \textbf{preference for heuristics over vocabulary-based design}. For example, participants who started with heuristic-guided sketching often intentionally applied the multi-layered heuristic when sketching vocabulary-based designs. However, we did not observe a tendency to re-apply visual vocabulary to heuristic-based tasks. T13 noted, \textit{``I felt freer to move beyond ellipses and boxes and to organize my ideas in a more trackable way.''}

\section{Evaluation}\label{sec:eval}

\begin{figure*}[t]
	\centering
  		\includegraphics[width=\linewidth]{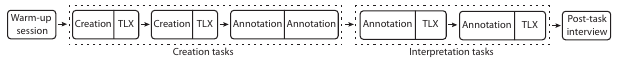}
		\caption{Each evaluation study included a warm-up session, two creation tasks (followed by an annotation process of privacy-related design decisions), two interpretation tasks, and a post-task interview. During annotation, participants also orally explained the content of each design decision. They completed a NASA TLX survey after each task.}
		\label{fig:study-flow}
\end{figure*}

This section presents a detailed experimental evaluation of the heuristic-based approach. Our results indicate that this approach helps data science students create high-quality privacy sketches quickly with reduced mental effort. Participants also reported that the guided privacy sketches were more readable than the baseline sketches.

\subsection{Study Overview}\label{eval:study}

\sssec{Participants.} \change{We recruited 24 participants in the evaluation through the same approach as previous studies (Sections~\ref{sec:needfing-method}, \ref{sec:teach_iter}).} Our final group included 10 undergraduates and 14 graduate students. \change{In the pre-screening privacy knowledge survey (same as Section~\ref{sec:needfing-method}), 14 participants answered ``No'' to the privacy knowledge question; six gave responses unrelated to privacy, and only four provided relevant privacy design terms but lacked experience in privacy design.}

Among our participants, 10 (41.7\%) identified as female, and others identified as male. Twenty (83.3\%) participants were aged between 18 and 24, while others were between 25 and 30 (Mean = 22.0 years, SD = 2.3 years). Each participant received a US \$15 Amazon gift card as compensation.

\sssec{Study Procedure and Apparatus.} We followed a between-subjects study design. Participants received either vocabulary-based or heuristic-based guidance for their tasks. 
They completed tasks on an iPad with a stylus using a sketch board application (e.g., Notability\change{~\cite{Notability}}).  

Figure~\ref{fig:study-flow} illustrates the study procedure. Each study began with a warm-up session (20-minute limit) that included an explanation of the worksheet (Table~\ref{tab:three-principle} or \ref{tab:attempt2-worksheet}) and sketching privacy design for a sample scenario \textit{Online Meeting Attention Tracking} (Table~\ref{tab:scenarios}). Researchers observed the process and answered questions as needed. 

Then, the study continued with two creation tasks in which participants sketched designs for two scenarios (i.e., \textit{Afterlife Chatbot} and \textit{Financial Risk Management}, as shown in Table~\ref{tab:scenarios}) and annotated their privacy design decisions in the sketches. The participants were randomly assigned to a scenario (Table~\ref{tab:create_order}), and upon completion, they would complete another. Each task has a 15-minute time limit. These time limits were decided based on the task completion time in the last experiment (Section~\ref{sec:attmpt3-find}).

Then, each participant was asked to interpret two sketches created by previous participants \change{(Table~\ref{tab:interp_order})}. Finally, we interviewed participants about what hindered or assisted them in their creation and interpretation tasks. 

\change{We took notes during each study session and summarized the key insights after each session. We observed saturation (i.e., no new insights emerged)~\cite{saunders2018saturation} after the twentieth study. We then stopped participant recruitment and concluded the evaluation study with four more participants\cite{francis2010adequate}.} On average, the study session lasted 90 minutes. 

\change{During analysis, we performed Mann-Whitney U tests~\cite{mcknight2010mann} to compare different scenario orders within each condition and found no evidence of ordering bias across all measures reported in Sections~\ref{sec:quantitative} and \ref{sec:interp-quant}. For example, within the vocabulary group, we assessed the first creation task's mental demand scores between participants who began with scenario F and those who began with scenario A.}

In the following sections, we will denote the Vocabulary-Based Sketching as ``Vocabulary'' and the Heuristic-Based Sketching as ``Heuristic.''

\sssec{\change{Annotating Visual Sketches.}}
Comparing high-dimensional sketches is challenging. To address this, we developed a codebook (Appendix Table~\ref{tab:design_hints}) to annotate privacy-related design decisions covered in the sketches. We then used the annotated decisions to assess the design coverage in sketch creation and the communication effectiveness between the sketch creators and interpreters \change{(Sections~\ref{sec:create-results}, \ref{sec:eval-interp})}.

\change{Since participants often cannot broadly explore the design space (Section~\ref{sec:needfing-method}), we cannot derive the potential design decisions entirely based on empirical observations. Instead, we began by collecting privacy-related design decisions from prior literature~\cite{solove2005taxonomy, jin2021lean, nist2019pram, al2008information, feng2021design, schaub2015design} to systematically capture the unique design decisions reflected in the sketches.} 
Two researchers then collaboratively created an initial codebook for analysis. They then independently applied the codebook to annotate the sketches \change{from previous studies (Sections~\ref{sec:needfing-method}, \ref{sec:teach_iter}) and discussed potential modifications, deletions, and extensions to the coding scheme. After refining the scheme, all these previous sketches} were re-coded using the updated framework. Given the subjective nature of interpreting sketches, we considered only codes independently validated by both researchers to ensure reliability.

\subsection{Creation Tasks}\label{sec:create-results}
\subsubsection{Data Collection}\label{sec:create-collect} We counterbalanced the presentation order of scenarios (Table~\ref{tab:create_order}). In each task, we asked participants to read the scenario description (Table~\ref{tab:scenarios}) and sketch a privacy design. They could inform us if they felt satisfied with their sketch, i.e., having included as many privacy-related details as possible, before the time was up. We recorded their completion time and asked them to fill out a NASA TLX survey~\cite{hart1988development} after each creation. 

We presented participants with a list of privacy-related design decisions (Table~\ref{tab:design_hints}) and asked them to describe each decision they had included, annotating the corresponding content on their sketches. We didn't set time limits for this process. Participants could skip any decisions they felt were not covered but could not modify their original designs.

\begin{table}[t!]
\captionof{table}{Creation task schedule. We counterbalanced the presentation order of the task scenarios (F: \textit{Financial Risk Management}; A: \textit{Afterlife Chatbot}) for both conditions (Vocabulary-guided and Heuristic-guided sketching).}
\centering
\resizebox{\linewidth}{!}{%
\begin{tabular}{|>{\centering\arraybackslash}p{1.5cm}|cc|cccccc|}
\hline
\multirow{2}{*}{Condition} & \multicolumn{2}{c|}{Scenario} & \multicolumn{6}{c|}{\multirow{2}{*}{Creator}} \\
                           & 1st & 2nd & \multicolumn{6}{c|}{}         \\ \hline\hline
\multirow{2}{*}{Heuristic} & F & A & P1 & P3  & P5  & P13 & P15 & P17 \\
                               & A & F & P7 & P9  & P11 & P19 & P21 & P23 \\ \hline
\multirow{2}{*}{Vocabulary} & F & A & P2 & P4  & P6  & P14 & P16 & P18 \\
                                    & A & F & P8 & P10 & P12 & P20 & P22 & P24 \\ \hline
\end{tabular}}
\label{tab:create_order}
\end{table}

\subsubsection{Data Analysis}\label{sec:create-analysis} \change{There is no perfect sketch for the creation task, as participants may have different designs for the same scenario, which may evolve during sketching. So, we use the codebook (Table~\ref{tab:design_hints}) to systematically capture the unique design decisions reflected in each sketch.} To assess the quality of the sketches, we documented the privacy design decisions annotated by participants, counted the \textbf{total number} of decisions included in each sketch, and calculated the \textbf{coverage percentage} of each design decision across all sketches in each condition.

Next, we conducted a thematic analysis~\cite{saldana2021coding} of post-task interview transcripts\change{~and notes we took during the study}. First, two researchers independently coded eight transcripts (four from the heuristic group and four from the vocabulary group). After discussing and incorporating codes, we created a codebook that the two researchers agreed on. Using this codebook, the researchers divided the remaining transcripts, each coding eight transcripts per condition. Like many other qualitative analyses in S\&P research~\cite{holtervennhoff2023wouldn, liu2024exploring}, the two researchers discussed and resolved coding conflicts in several weekly meetings. In this analysis, as we prioritized identifying emerging themes, we did not calculate the inter-rater reliability (IRR) to seek theoretical agreement~\cite{mcdonald2019reliability}.

\begin{table}[t]
\centering
\caption{In creation tasks, heuristic-guided participants spent less time sketching until satisfaction and included more privacy-related design decisions in their sketches. Format: mean $\pm$ standard deviation. \change{We also highlight the higher value between two conditions.}}
\resizebox{\linewidth}{!}{%
\begin{tabular}{|c|cc|cc|}
\hline
\multicolumn{1}{|c|}{\multirow{2}{*}{Task Order }}  & \multicolumn{2}{c|}{Vocabulary} & \multicolumn{2}{c|}{Heuristic} \\ 
\multicolumn{1}{|c|}{}  & Time (min)         & \# Decisions         & Time (min)         & \# Decisions           \\ \hline
First       &   14.89 $\pm$ 2.86   &  8.4 $\pm$ 3.0       &    \textbf{12.70} $\pm$ 0.86   &     \textbf{10.7} $\pm$ 2.8   \\
Second      &    14.70 $\pm$ 3.71   &   9.3 $\pm$ 3.1       &    \textbf{10.36} $\pm$ 0.98  &    \textbf{12.3} $\pm$ 1.8     \\
All         &   14.80  $\pm$ 3.23    &    8.9 $\pm$ 3.0     &    \textbf{11.53} $\pm$ 0.94    &  \textbf{11.5}  $\pm$ 2.4    \\\hline

\end{tabular}\label{create-time-nele}
}
\end{table}

\subsubsection{Quantitative Results}\label{sec:quantitative}

\begin{table*}[t]

\caption{In the second creation task, heuristic-guided participants reported more positive responses than they did in the first task, a trend not observed in the vocabulary group. We present NASA TLX results (scale of 1 to 5) as ``median (mean $\pm$ standard deviation),'' and highlight second task responses if they are more positive than the first. ``$^{*}$'' indicates statistical significance ($p~\textless~0.05$) under Wilcoxon Signed-Rank test.\change{~``$\downarrow$'' denotes that a lower value is a more positive outcome.}}
\centering
\resizebox{\textwidth}{!}{%
\begin{tabular}{|cc|cccccc|}
\hline
 Condition & Task Order & Mental Demand~$\downarrow$ & Physical Demand~$\downarrow$ & Temporal Demand~$\downarrow$ & Performance~$\uparrow$ & Effort~$\downarrow$& Frustration~$\downarrow$\\
\hline
\multirow{2}{*}{Vocabulary} & First & 3.5 (3.17 $\pm$ 1.03) & 4.0 (3.65 $\pm$ 1.23) & 4.0 (3.33 $\pm$ 0.98) & 3.0 (3.00 $\pm$ 0.74) & 4.0 (3.83 $\pm$ 0.39)& 2.0 (2.50 $\pm$ 0.90) \\
 & Second & 3.5 (3.33 $\pm$ 0.78) & 4.0 (3.81 $\pm$ 1.31) & 3.5 (\textbf{3.17} $\pm$ 0.94) & 3.0 (2.92 $\pm$ 1.08) &3.5 (\textbf{3.50} $\pm$ 0.80)& 3.0 (2.67 $\pm$ 0.89)\\
\hline
\multirow{2}{*}{Heuristic} & First & 4.0 (3.67 $\pm$ 0.89)$^{*}$ & 3.5 (3.48 $\pm$ 0.97) & 4.0 (3.58 $\pm$ 0.79) & 3.0 (2.92 $\pm$ 0.67) & 3.5 (3.58 $\pm$ 0.67) &3.0 (3.00 $\pm$ 1.04)\\
 & Second & 2.5 (\textbf{2.92} $\pm$ 1.24)$^{*}$  & 3.0 (\textbf{2.97} $\pm$ 0.82) & 3.0 (\textbf{3.00} $\pm$ 1.41) & 3.5 (\textbf{3.25} $\pm$ 0.87) & 3.0 (\textbf{3.17} $\pm$ 1.19)& 2.0 (\textbf{2.33} $\pm$ 1.07) \\
\hline
\end{tabular}
} 
\label{tab:creation_tlx}
\end{table*}

We found that heuristics could help students sketch privacy designs more effectively. Their designs also demonstrated higher quality.

\sssec{Improved Sketching Efficiency.} As shown in Table~\ref{create-time-nele}, heuristic-guided participants reached their satisfied designs faster (average of 11.53 vs. 14.80 minutes) and covered more privacy-related design decisions (average of 11.5 vs. 8.9 per sketch), compared to the vocabulary group.

Besides, as they progressed to the second creation task, the heuristic-guided participants reported more positive responses to the NASA TLX questions (Table~\ref{tab:creation_tlx}) than the first task. These improvements included a significantly reduced mental workload shown by the Wilcoxon Signed-Rank test~\cite{woolson2005wilcoxon} ($p~\textless~0.05\change{, r=0.70}$), as well as decreases in physical, temporal, and effort demands. The heuristic group also noted increased self-perceived performance and reduced frustration, a trend not observed in the vocabulary group.

\sssec{Broader Coverage of Privacy-Related Design Decisions.} We then compared the coverage percentage of design decisions between conditions (Table~\ref{tab:ele_cover_rate}). On average, each decision was included in 76.7\% of the heuristic-guided sketches, compared to 59.2\% of the vocabulary group. Specifically, heuristic group could more frequently cover decisions such as \textit{Stored Data}, \textit{Processing Input}, \textit{Processing Approach}, \textit{Accessed Output}, \textit{Access Approach}, and \textit{Choice Impacts}. This difference was statistically significant ($p~\textless~0.01\change{, r=0.69}$) under the Wilcoxon Signed-Rank test~\cite{woolson2005wilcoxon}.

\subsubsection{Qualitative Findings}\label{sec:creat-qual}
Based on the analysis of participants' creation task behaviors and their feedback in post-task interviews, we identified following main findings.

\sssec{Reduced Learning Curve.} The guidelines provide a starting point and ease the challenge of beginning a sketch from scratch. For example, P15 shared that \textit{``The heuristic helps me to know what the design should roughly look like when I started to construct my ideas.''} P7 added that \textit{``Including stakeholders prompts me to consider who is involved and who is taking action in data cases, which I believe is important when forming my ideas for sketches.''}

\sssec{Flexibility in Planning.} Participants appreciate the \textbf{Multi-layered Representation} heuristic for helping them in planning their sketches in a structured manner. P7 explained, \textit{``The overview layer allows me to lay down the general ideas so I can worry about the details later.''} The device-based data flow and multi-layered approach also gave them a \textit{``big-picture perspective''} (P3), which helped P5 to \textit{``organize the system’s overall logic and focus on how data flows between different modules without getting distracted by other details.''} Participants also highlighted that the separate layers enable them to \textit{``quickly update on details''} (P9) and \textit{``easily to replicate existing content''} (P11). Furthermore, as they were familiar with the heuristics after the first design, participants in the experimental group were able to quickly grasp and re-apply the workflow to the second task. 

\sssec{Ease of Cognitive Load.} After applying the heuristics, participants recognized that sketching was more manageable than anticipated, reducing their perceived effort and workload. P7 expressed that using the device annotation made \textit{``each part more distinctive and save effort if I want to adding new ideas to my sketch.''} P19 initially worried about \textit{``how to show all the privacy details''} but later found that \textit{``the layered approach helped me break down complex ideas into simple parts.''} Despite the initial time pressure, P5 noted the benefit of sketching the overview layer first, explaining that it allows for \textit{``managing time more effectively by following the outline I planned out from the start.''}

\sssec{\change{Enhanced Coverage of Privacy Details.}} \change{The heuristics prompt participants to extend the scope of privacy details in their sketches, echoing the increased design decision coverage shown in Table \ref{tab:ele_cover_rate}. First, the \textbf{Stakeholder Interactions with Data Flow} heuristic encouraged the participants to \textit{``consider human decisions and accountability with freedom''} in a data practice (P11), which is \textit{``important for ensuring responsible use of data''} (P7). P23 explained that the heuristic enabled thinking \textit{``beyond where the data travels in the system''} and encouraged consideration of \textit{``what each party is doing with the data.''}} 

\change{Second, the \textbf{Multi-Layered Representation} heuristic encourages more detailed privacy designs. Participants noted that the separation of layers \textit{``makes space for me to fill in details''} (P9) and that the flexibility to update individual layers enabled easy extension of their sketches, encouraging them \textit{``to iterate more times without friction''} (P3).} 

\change{As a result, compared to the vocabulary group, heuristic-guided sketches incorporated more interaction details beyond merely including “Choice” and “Involvement.” For example, as illustrated in Figure~\ref{fig:title_fig}, these sketches featured interactive elements such as a chat window for access or a pop-up for making choices. This corresponded to the increased coverage (Table \ref{tab:ele_cover_rate}) of the three decisions related to \textit{Choice \& Notice}—\textit{Choice Options}, \textit{Choice Impacts}, and \textit{Choice Notification}.}

\sssec{Bias and Negative Consequence.} Participants noted certain limitations with the heuristics during the sketching process. P1 remarked that the Multi-layered Representation \textit{``seems to be repetitive because I am filling similar information in both layers.''} P7 observed that creating a sketch on the given case was \textit{``a systematic design''} and expressed a desire for a more \textit{``complete structured framework to fill or follow.''}

\begin{table*}[t]
\centering
\caption{Heuristic-guided sketches could include privacy design decisions more frequently. Additionally, the design decisions within the heuristic-guided sketches were interpreted more accurately than those within the vocabulary group. For each design decision, we report its coverage frequency (``Coverage'') across all sketches in each group. For communication, we report the frequency of being interpreted (``Interpret,'' including misinterpretations), along with precision, recall, and F1-score, all measured at the design decision level. For each metric, we \change{highlight the higher value and~}indicate the statistical significance between the two conditions using the Wilcoxon Signed-Rank tests ($^{*}p~\textless~0.05$, $^{**}p~\textless~0.01$, $^{***}p~\textless~0.001$)\change{, all showing large effect sizes ($r~\textgreater~0.5$)}.}
\resizebox{.992\textwidth}{!}{%
\begin{tabular}{|l|>{\centering\arraybackslash}p{1.0cm}>{\centering\arraybackslash}p{1.0cm}>{\centering\arraybackslash}p{1.1cm}>{\centering\arraybackslash}p{.8cm}>{\centering\arraybackslash}p{.8cm}|>{\centering\arraybackslash}p{1.0cm}>{\centering\arraybackslash}p{1.0cm}>{\centering\arraybackslash}p{1.1cm}>{\centering\arraybackslash}p{.8cm}>{\centering\arraybackslash}p{.8cm}|}\hline
\multicolumn{1}{|c|}{\multirow{2}{*}{Design Decision}} & \multicolumn{5}{c|}{Vocabulary-Guided Sketches}                   & \multicolumn{5}{c|}{Heuristic-Guided Sketches}              \\
\multicolumn{1}{|c|}{}                         & Coverage & Interpret & Precision & Recall & F1 & Coverage & Interpret & Precision & Recall & F1 \\ \hline\hline
Collected Personal Data & \multicolumn{1}{c|}{\textbf{100.0\%}} & 86.4\% & 83.3\% & 92.1\% & 87.5\% & \multicolumn{1}{c|}{95.8\%}  & \textbf{87.0\%} & \textbf{90.9\%} & \textbf{100.0\%} & \textbf{95.2\%} \\
Data Provider           & \multicolumn{1}{c|}{87.5\%}  & 95.5\% & 70.5\% & 73.8\% & 72.1\% & \multicolumn{1}{c|}{\textbf{91.7\%}}  & \textbf{95.7\%} &\textbf{ 91.3\%} & \textbf{95.5\%}  & \textbf{93.3\%} \\
Collection Purpose      & \multicolumn{1}{c|}{\textbf{87.5\%}}  & \textbf{77.3\%} & 27.8\% & 29.4\% & 28.6\% & \multicolumn{1}{c|}{83.3\%}  & 73.9\% & \textbf{65.0\%} & \textbf{76.5\%}  & \textbf{70.3\%} \\
Stored Data    & \multicolumn{1}{c|}{62.5\%}  & 77.3\% & 34.4\% & 32.4\% & 33.3\% & \multicolumn{1}{c|}{\textbf{91.7\%}}  & \textbf{78.3\%} & \textbf{77.8\%} & \textbf{77.8\%}  & \textbf{77.8\%} \\
Storage Approach        & \multicolumn{1}{c|}{79.2\%}  & 63.6\% & 33.3\% & 42.9\% & 37.5\% & \multicolumn{1}{c|}{\textbf{91.7\%}}  & \textbf{73.9\%} & \textbf{78.9\%} & \textbf{88.2\%}  & \textbf{83.3\%} \\
Post-Storage Action     & \multicolumn{1}{c|}{58.3\%}  & 31.8\% & 10.0\% & 21.4\% & 13.6\% & \multicolumn{1}{c|}{\textbf{62.5\%}}  & \textbf{60.9\%} & \textbf{56.3\%} & \textbf{64.3\%}  & \textbf{60.0\%} \\
Processed Input & \multicolumn{1}{c|}{54.2\%}  & \textbf{81.8\%} & 29.4\% & 27.8\% & 28.6\% & \multicolumn{1}{c|}{\textbf{100.0\%}} & 78.3\% & \textbf{67.5\%} & \textbf{75.0\%}  & \textbf{71.1\%} \\
Processing Output       & \multicolumn{1}{c|}{79.2\%}  & \textbf{81.8\%} & 55.3\% & 58.3\% & 56.8\% & \multicolumn{1}{c|}{\textbf{91.7\%}}  & 65.2\% & \textbf{85.3\%} & \textbf{96.7\%}  & \textbf{90.6\%} \\
Processing Approach     & \multicolumn{1}{c|}{45.8\%}  & 50.0\% & 29.2\% & 31.8\% & 30.4\% & \multicolumn{1}{c|}{\textbf{83.3\%}}  & \textbf{60.9\%} & \textbf{66.7\%} & \textbf{71.4\%}  & \textbf{69.0\%} \\
Accessed Raw Data  & \multicolumn{1}{c|}{\textbf{66.7\%}}  & \textbf{54.5\%} & 21.4\% & 25.0\% & 23.1\% & \multicolumn{1}{c|}{50.0\%}  & 52.2\% & \textbf{59.1\%} & \textbf{54.2\%}  & \textbf{56.5\%} \\
Accessed Output                   & \multicolumn{1}{c|}{41.7\%}   & \textbf{63.6\%}   & \textbf{58.3\%}    & 50.0\% & 53.8\%   & \multicolumn{1}{c|}{\textbf{75.0\%}}   & 60.9\%   & 52.8\%    & \textbf{67.9\%} & \textbf{59.4\%}   \\
Access Approach         & \multicolumn{1}{c|}{50.0\%}  & 59.1\% & 31.8\% & 26.9\% & 29.2\% & \multicolumn{1}{c|}{\textbf{83.3\%}}  & \textbf{69.6\%} & \textbf{66.7\%} & \textbf{62.5\%} & \textbf{64.5\%} \\
Choice Options      & \multicolumn{1}{c|}{37.5\%}  & 50.0\% & 45.5\% & 45.5\% & 45.5\% & \multicolumn{1}{c|}{\textbf{58.3\%}}  & \textbf{52.2\%} & \textbf{64.3\%} & \textbf{75.0\%}  & \textbf{69.2\%} \\
Choice Impacts       & \multicolumn{1}{c|}{16.7\%}  & \textbf{36.4\%} & 14.3\% & 12.5\% & 13.3\% & \multicolumn{1}{c|}{\textbf{58.3\%}}  & 30.4\% & \textbf{37.5\%} & \textbf{64.3\%}  & \textbf{47.4\%} \\
Choice Notification    & \multicolumn{1}{c|}{20.8\%}  & 18.2\% & 16.7\% & 25.0\% & 20.0\% & \multicolumn{1}{c|}{\textbf{33.3\%}}  & \textbf{26.1\%} & \textbf{75.0\%} & \textbf{75.0\%}  & \textbf{75.0\%} \\ \hline\hline
\multicolumn{1}{|c|}{\multirow{1}{*}{Mean}}    & \multicolumn{1}{c|}{59.2\%}  & 61.8\% & 37.4\% & 39.7\% & 38.2\% & \multicolumn{1}{c|}{\textbf{76.7\%}}  & \textbf{64.4\%} & \textbf{69.0\%} & \textbf{76.3\%}  & \textbf{72.2\%} \\ 
\multicolumn{1}{|c|}{\multirow{1}{*}{$p$-value}}    & \multicolumn{1}{c|}{$^{**}$}  & $(n.s.)$ & $^{***}$ & $^{***}$ & $^{***}$ & \multicolumn{1}{c|}{$^{**}$}  & $(n.s.)$ & $^{***}$ & $^{***}$ & $^{***}$ \\ \hline

\end{tabular}
}
\label{tab:ele_cover_rate}
\end{table*}

\subsection{Interpretation Tasks}\label{sec:eval-interp}

\begin{table}[h!]
\captionof{table}{Interpretation task schedule. For instance, P3 first interpreted a sketch of \textit{Financial Risk Management}, which was created by P1 (heuristic group), and then interpreted a sketch of another scenario by P2 (vocabulary group). The ``-'' means no interpretation task due to the lack of a prior sketch, and three sketches (one by P23 and two by P24) were not interpreted because of no subsequent interpreters.}
\centering
\resizebox{\linewidth}{!}{%
\begin{tabular}{|>{\centering\arraybackslash}p{1.14cm}|>{\centering\arraybackslash}p{.72cm}>{\centering\arraybackslash}p{.72cm}|>{\centering\arraybackslash}p{1.14cm}||>{\centering\arraybackslash}p{1.14cm}|>{\centering\arraybackslash}p{.72cm}>{\centering\arraybackslash}p{.72cm}|>{\centering\arraybackslash}p{1.14cm}|}
\hline
Scenario & F & A & \multirow{2}{*}{Interpreter} & Scenario & A & F & \multirow{2}{*}{Interpreter} \\
Condition & Heuris. & Vocab. &  & Condition & Vocab. & Heuris. &  \\ \hline\hline
\multirow{6}{*}{Creator} & - & - & P1 & \multirow{6}{*}{Creator} & P6 & P5 & P7 \\
                         & P1 & P2 & P3 &  & P8 & P7 & P9 \\
                         & P3 & P4 & P5 &  & P10 & P9 & P11 \\
                         & P11 & P12 & P13 &  & P18 & P17 & P19 \\
                         & P13 & P14 & P15 &  & P20 & P19 & P21 \\
                         & P15 & P16 & P17 &  & P22 & P21 & P23 \\ \hline\hline
Scenario & F & A & \multirow{2}{*}{Interpreter} & Scenario & A & F & \multirow{2}{*}{Interpreter} \\
Condition & Vocab. & Heuris. &  & Condition & Heuris. & Vocab. &  \\ \hline\hline
\multirow{6}{*}{Creator} & - & P1 & P2 & \multirow{6}{*}{Creator} & P7 & P6 & P8 \\
                         & P2 & P3 & P4 &  & P9 & P8 & P10 \\
                         & P4 & P5 & P6 &  & P11 & P10 & P12 \\
                         & P12 & P13 & P14 &  & P19 & P18 & P20 \\
                         & P14 & P15 & P16 &  & P21 & P20 & P22 \\
                         & P16 & P17 & P18 &  & P23 & P22 & P24 \\ \hline
\end{tabular}
}
\label{tab:interp_order}
\end{table}

\subsubsection{Data Collection}\label{sec:interp-method} As shown in Figure~\ref{fig:study-flow}, each participant completed two interpretation tasks, each with a 7-minute time limit. In each task, they received a randomized list of privacy-related design decisions (excluding the ``Procedure'' column in Table~\ref{tab:design_hints}) and an unannotated copy of a sketch created by a prior participant. We then asked them to annotate and orally describe the privacy-related design decisions they identified. This annotation is similar to what they have done during the creation tasks (Section~\ref{sec:create-collect}). Table~\ref{tab:interp_order} outlines the interpretation task order, where each participant interpreted two sketches of different scenarios: one created by a heuristic-guided participant and another from the vocabulary group. We maintained scenario order consistent with the creation tasks (Table~\ref{tab:create_order}) to avoid biases from the most recent scenario they had sketched. For example, after sketching for \textit{Afterlife Chatbot} as his second creation, P3 would first interpret a sketch of another scenario. After each task, we recorded their interpretation time and asked them to complete a NASA TLX survey~\cite{hart1988development}.

\subsubsection{Data Analysis}\label{sec:interp-method-analysis}

To assess whether heuristic-guided sketches could better facilitate communication of privacy design, we measured the alignment between the creator's and interpreter's responses (including annotations and explanations) regarding each privacy-related design decision within a sketch. For each design decision, two researchers collaboratively compared both response versions and assigned a code to describe whether they aligned. They open-coded a subset of sketches (N=10, i.e., 15 codes per sketch $\times$ 10 sketches), conducted three rounds of discussions to reach a complete agreement, and generated an initial codebook. Then, two researchers coded the rest of the sketches independently. To ensure consistency, two coders discussed their codes regularly to reach an agreement and iteratively refined the codebook (e.g., an initial code ``only one person gave a response'' was decoupled into two codes ``only the creator responded'' and ``only interpreter responded''). Finally, two researchers coded all the sketches using the final codebook (Appendix Table~\ref{tab:six-pattern}), with a Cohen's kappa~\cite{cohen1960coefficient} of 90.1\%, which reflects an ``almost perfect agreement''~\cite{mchugh2012interrater}.

We then quantified the effectiveness of communication on two levels. First, to measure whether heuristic-guided sketches lead to high-quality interpretations for each privacy-related \textit{design decision}, we used the creators' responses as groundtruth and calculated \textit{design decision-level} \textbf{precision} (i.e., the proportion of interpretations that matched the groundtruth) as well as \textbf{recall} (i.e., the proportion of groundtruth responses that were accurately interpreted).

Second, to determine whether sketching with heuristics aids in identifying correct privacy design decisions across \textit{an entire sketch}, we computed \textit{sketch-level} \textbf{precision} (the proportion of interpretations that matched the creator's responses within each sketch) and \textbf{recall} (the proportion of design decisions covered by the creator that was accurately interpreted in each sketch). Appendix Table~\ref{apdx:confusion-mat} provides further clarification of these metrics.

The qualitative analysis shared the same procedure as described in Section~\ref{sec:create-analysis}, and we will report the findings of interpretation tasks in Section~\ref{sec:interp-qual}.

\begin{table*}[t]
\caption{Participants perceived significantly lower workloads when interpreting heuristic-guided sketches than vocabulary-guided ones. We present their responses to NASA TLX survey as median (mean $\pm$ standard deviation), with results for heuristic-guided sketches highlighted. We annotate statistically significant improvements based on the Wilcoxon Signed-Rank tests ($^{*}p~\textless~0.05$, $^{**}p~\textless~0.01$, $^{***}p~\textless~0.001$)\change{, all with medium to near large effect sizes ($r: 0.3\sim0.5$)}.\change{~``$\downarrow$'' indicates that a lower value is more positive.}}
\centering
\resizebox{\textwidth}{!}{%
\begin{tabular}{|c|cccccc|}
\hline
 Condition  & Mental Demand~$\downarrow$ & Physical Demand~$\downarrow$ & Temporal Demand~$\downarrow$ & Performance~$\uparrow$ & Effort~$\downarrow$& Frustration~$\downarrow$\\
\hline
Vocabulary & 3.0 (3.41 $\pm$ 1.10)$^{***}$ & 3.5 (3.66 $\pm$ 0.92)$^{**}$ & 3.0 (2.82 $\pm$ 1.18)$^{*}$ & 3.0 (2.91 $\pm$ 1.15)$^{**}$ & 3.5 (3.32 $\pm$ 1.04)$^{***}$& 3.0 (3.23 $\pm$ 1.07)$^{***}$ \\
Heuristic & \textbf{2.0 (2.29 $\pm$ 0.86)$^{***}$} & \textbf{2.5 (2.61 $\pm$ 0.88)$^{**}$} & \textbf{2.0 (2.17 $\pm$ 0.87)$^{*}$} & \textbf{4.0 (3.75 $\pm$ 0.79)$^{**}$} &\textbf{2.0 (2.25 $\pm$ 0.85)$^{***}$}& \textbf{2.0 (2.04 $\pm$ 0.81)$^{***}$} \\
\hline
\end{tabular}}
\label{tab:interp_tlx}
\end{table*}

\subsubsection{Quantitative Results}\label{sec:interp-quant} We found that heuristic-guided sketches are easier to interpret and could better facilitate communication between creators and interpreters.

\sssec{Eased Workload of Interpretation.} All participants finished their interpretation within the time limit. \change{There was no significant difference in interpretation time (4.9 minutes for heuristic-guided sketches vs. 5.1 minutes for vocabulary-guided sketches).} However, participants consistently reported significantly lower mental, physical, temporal, and effort-related demands and reduced frustration when working with heuristic-guided sketches. They also perceived improved performance, as indicated by Wilcoxon Signed-Rank tests (Table~\ref{tab:interp_tlx}). Furthermore, participants identified more privacy design decisions in heuristic-guided sketches, with an average of 9.4 elements per sketch compared to 8.1 from vocabulary-guided ones. These results suggest that heuristic-guided sketches are more straightforward to interpret, decreasing workload while enhancing outcomes.

\sssec{Improved Communication Efficiency.} Table~\ref{tab:ele_cover_rate} presents the result of interpretation evaluation metrics at the privacy design decision level, which provides initial evidence for heuristic guidance's capability to enable better creator-interpreter communication of privacy designs. First, heuristic-guided sketches achieved higher interpretation performance across metrics, with an increase of 31.6\% in precision (i.e., increased fraction of correct interpretations) and 36.6\% in recall (i.e., increased fraction of creator's design decisions that were accurately interpreted). 

At the level of a complete privacy design, heuristic-guided sketches demonstrated higher per-sketch interpretation performance, again with a higher precision (79.1\% vs. 44.6\% for the vocabulary group), reflecting the proportion of matches among all interpretations in a sketch, and higher recall (71.1\% vs. 40.5\%), representing the percentage of alignments among the creator's all design decisions. These differences are all statistically significant according to the Wilcoxon Signed-Rank tests~\cite{woolson2005wilcoxon}, highlighting the improved communication efficiency facilitated by heuristic guidance.

\begin{figure}[t!]
	\centering
  		\includegraphics[width=\linewidth]{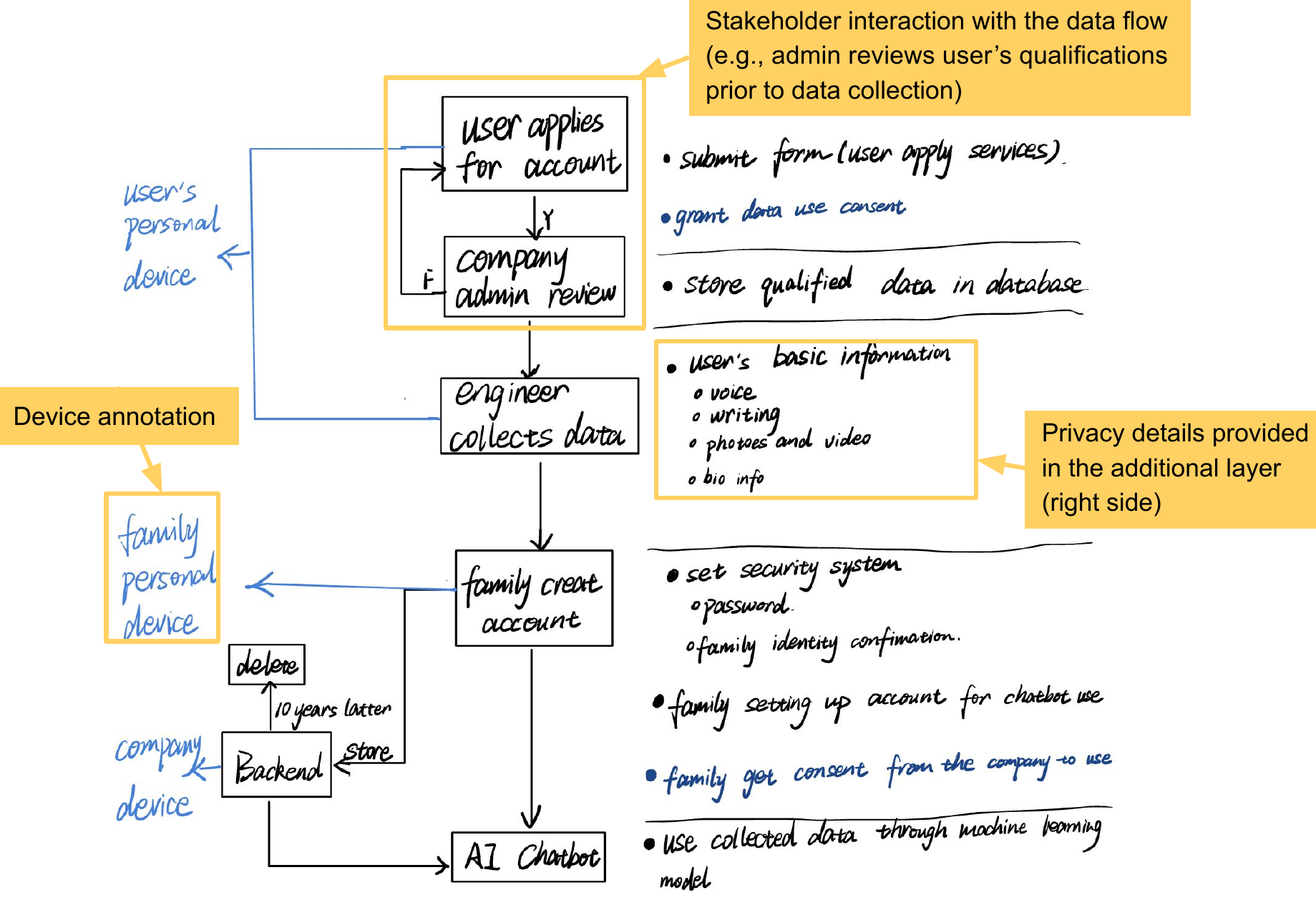}
		\caption{A heuristic-guided privacy design sketch for the \textit{Afterlife Chatbot} scenario. It presents privacy details in parallel with the corresponding devices, stakeholders, and the steps taken, offering a clear logic for readers to follow.}
		\label{fig:discuss-fig}
\end{figure}

\subsubsection{Qualitative Findings}\label{sec:interp-qual}
We made the following findings based on participants' behaviors and feedback about their interpretation tasks.

\sssec{Heuristics streamline the interpretation process.} \change{Participants from both conditions found heuristic-guided sketches to be more structured and easier to understand, largely due to the enhanced readability provided by the \textbf{Multi-Layered Representation} heuristic. P17, a heuristic-guided participant, appreciated how the multi-layered design allowed him to quickly grasp the overall idea of the privacy design at first glance, noting that \textit{``[reading] this sketch felt quite similar to [creating] my own sketch work.''}} 

\change{Similarly, P10, who did not receive heuristic instruction, also found the sketch (Figure~\ref{fig:discuss-fig}) easier to follow, explaining that \textit{``the overview layer explains a clear logic of the design, such as where the privacy-sensitive procedure happens and who were involved in. However, the other [vocabulary-guided] sketch obviously falls short in this clarity.''} She further expressed interest in \textit{``adopt[ing] this layered style in my future design sketches.''} P4 also emphasized that separating the overview from details \textit{``enables a progressive manner of reading,''} while P18 highlighted how \textit{``this multi-level sketch prevents my visual overload from digesting too much content in one place.''}}

\sssec{Heuristics offer varying content density.} Participants showed different preferences when retrieving information: some preferred more text for detailed understanding, while others favored less text for a smoother reading experience. The heuristic-guided sketches effectively accommodated both needs. P6 praised the multi-layered sketch for its support of effective information retrieval, saying, \textit{``first, I skim through the overview of the data flow to understand the big picture. If I want specific details, like the data involved in each step, I go to the next layer using the annotation labels.''} P10 described the device box and overview layer as \textit{``pretty concise and simple,''} while preferring more text for detailed understanding. P11 appreciated that the \textit{``detail layer has enough information for me to understand creator's thought,''} but felt that excessive text made interpretation time-consuming.

\section{Discussion \& Future Work}\label{sec:discussion}

\sssec{\change{Connections to Established Practices}}. 
\change{Professionals often draw on their existing skill sets when sketching for privacy. For instance, UX designers may use Customer Journey Maps, while software engineers rely on UML diagrams. 
We chose to develop a tailored method from scratch for two reasons. First, these skill adaptions only partially address key aspects of privacy design. For example, Customer Journey Maps focuses on stakeholder interactions. Second, most of our target audience—data science students—lacks formal training in both usability research and software engineering.}

\change{We have incorporated multiple relevant ideas in software engineering and UX research into our solutions, such as Data Flow Diagrams, Customer Journey Maps, and UX storyboards~\cite{truong2006storyboarding}. Future work may investigate whether our proposed heuristics can also enhance the practices of experienced professionals and how these heuristics complement their existing techniques.}

\sssec{Why do heuristics work in sketching privacy?} Unlike rigid frameworks that limit creativity and restrict exploration \cite{alhirabi2023parrot,LucidchartDFD}, the heuristic-based approach emphasizes flexibility and provides simple guidance, enabling students to deeply engage with privacy contexts while effectively organizing their ideas. Furthermore, the \textbf{Multi-layer Representation} heuristic helps students manage their sketches from a macro perspective, allowing them to approach designs strategically.

\sssec{Sketches vs. Diagrams.} We experimented with both sketches and diagrams in this project. We found that complex diagram structures hindered the learning process (Section~\ref{sec:attempt1-find}). For instance, when students used diagrams to illustrate interactions between stakeholders and data flows, they needed a thorough understanding of diagram vocabulary. This steep learning curve increased their cognitive load, making it harder to complete the privacy design.

In contrast, sketches mitigated these challenges by enabling more intuitive idea generation (Section~\ref{sec:create-results}) and communication (Section~\ref{sec:eval-interp}). This aligns with prior research highlighting sketching as a flexible and intuitive tool for ideation and communication \cite{landay1995interactive,kelley2017sketching}. However, sketches lack the structured detail of diagrams, which is crucial for tasks requiring precision and depth. While sketches reduce cognitive demands, they may limit students' ability to achieve the detailed understanding offered by structured diagrams. Future work could explore methods to help students effectively use diagrams to communicate privacy designs.

\sssec{Sketching for Privacy Literacy.} In this paper, our educational audience is data science students who are future data practitioners. One potential future direction is to push the technique to layperson~\cite{oates2018turtles}, and explore if sketching privacy can help them form a more precise mental model~\cite{norman2013design} of the data practices.

\sssec{High-fidelity Privacy Prototypes.} Two students with UX/UI backgrounds noted that, although they were generally satisfied with their privacy designs, they felt uneasy about their inability to fully depict the visual aspects of the product's interface. They also found that messy hand sketches contributed to difficulties in interpretation. Future work could explore higher-fidelity privacy prototypes, akin to high-fidelity UX prototypes.

\sssec{\change{Sketches and Threat Modeling.}} 
\change{
Our current study focuses on facilitating the communication of privacy design concepts through sketches. Future research may explore the benefits of sketches in other security and privacy applications. For example, participants may develop a stronger awareness of privacy risks during sketching~\cite{stevens2018battle, thompson2024there,van2022descriptive,shreeve2020so}.
The improved communication enabled by privacy sketches may enhance the collaborative decision-making process~\cite{degeling2016privacy}.
Conventional threat modeling approaches~\cite{shullEvaluationThreatModeling, shevchenko2018threat} often rely on static tables or Data Flow Diagrams (DFDs) to define technical scope and decompose applications. Compared to tables and DFDs, sketching may better capture dynamic aspects of a system, such as stakeholder interactions and contextual nuances that are difficult to convey textually.}

\section{Limitations}

\sssec{\change{Ground Truth for the Creation Task.}} \change{The creation task has no ground truth. While our codebook (Table~\ref{tab:design_hints}) provided a structured approach for labeling and interpreting data, it may have constrained the granularity of our analysis. By relying on predefined categories, we may have overlooked subtle contextual differences or nuances in participant responses.}

\sssec{Participant Pool Limitations.} 
Our participant pool consisted primarily of students from four-year U.S. universities, which introduced limitations in both sample diversity and size. In particular, some participants had prior exposure to privacy-related concepts, which may not reflect the perspectives of novice users who engage with privacy features with little to no foundational knowledge. 
Future research may expand recruitment to include participants from community colleges and technical bootcamps, capturing a broader range of educational backgrounds and user experiences.

\section{Conclusion}\label{sec:conclusion design}
This paper explores methods for teaching data science students to sketch privacy designs. Through a need-finding study (N=12), we identified three challenges students encounter when sketching privacy designs: (1) difficulty sketching a complete data flow, (2) omission of stakeholder interactions with the data flow, such as user consent, and (3) failure to make quick updates without disrupting the overall design. To address these challenges, we iteratively developed our teaching approach, ultimately leading to three heuristics: (1) Device Annotation, (2) Stakeholder Interaction with Data Flow, and (3) Multi-Layered Representation. Our between-subjects experiment (N=24) demonstrates the effectiveness of these heuristics in improving sketching quality and facilitating communication of privacy design. Future privacy education and privacy design toolkits could take our findings into consideration.

\section{Research Ethics}
This research received approval from the Institutional Review Board (IRB) of our institution. 
\change{We conducted all studies using Zoom (approved by our institute) and used Zoom's speech recognition feature to transcribe the audio. 
To enable the transcription, we had to turn on the audio recording. 
Participants were informed about Zoom’s transcription and recording features, and a pop-up notification appeared when these features were activated. After each interview, the research team checked the transcription using the recording, then immediately and permanently deleted the recording. As such, we did not store any audio recordings. 
In this process, we have removed all personal information (e.g., names) from the transcription, ensuring that none of the transcriptions are personally identifiable.} Participants provided informed consent before the study and retained the right to withdraw at any stage. Throughout all phases of data collection (Sections~\ref{sec:needfing-method}, \ref{sec:teach_iter}, \ref{sec:eval}), participants had the right to withdraw at any time.

\section*{\change{Acknowledgement}}
\change{We thank the anonymous reviewers and the shepherd for their invaluable feedback and the participants for their kind involvement in our studies. This research is in part supported by a gift from Solix Technologies, the National Science Foundation CNS-2341187, CNS-2426397, a Google PSS Faculty Award, and a Meta Research Award.}

\bibliographystyle{IEEEtran}
\balance
\bibliography{main}

\begingroup
\renewcommand{\section}[2]{}  

\endgroup

\onecolumn
\appendices 

\section{15 Privacy-Related Design Decisions}

\begin{table*}[h]
\centering
\caption{We developed a codebook of 15 privacy-related design decisions to assess the privacy-related design decisions covered in the sketches and the communication effectiveness between the sketch creator and interpreter.}
\resizebox{\linewidth}{!}{%
\begin{tabular}{|m{.11\linewidth}|l|l|m{.25\linewidth}|}
\hline
\multicolumn{1}{|c|}{\textbf{Procedure}} & \multicolumn{1}{c|}{\textbf{Design Decision}} & \multicolumn{1}{c|}{\textbf{Explanation}} & \multicolumn{1}{c|}{\textbf{Example}} \\ \hline\hline
\multirow{3}{*}{Data Collection} 
    & Collected Personal Data & Personal data collected to achieve goals & SSN, transaction history, photo, IP address \\ \cline{2-4} 
    & Data Provider           & Individuals from whom data is collected & User, attendee \\ \cline{2-4} 
    & Collection Purpose      & Purpose for data collection & Prediction, decision making \\ \hline
\multirow{3}{*}{\vspace{-.35cm}Data Retention}   
    & Stored Data            & Personal data being stored & Transaction history, photo \\ \cline{2-4} 
    & Storage Approach       & Data storage implementation details & Data stored on cloud server for 3 months \\ \cline{2-4} 
    & Post-Storage Action    & Action taken after current data storage ends & Data transferred to an archive server after being deleted on the main server \\ \hline
\multirow{3}{*}{\vspace{-.5cm}Data Processing}  
    & Processing Input       & Personal data to be processed & Transaction history, photo \\ \cline{2-4} 
    & Processing Output      & Output data after processing & Risk score derived from user's history, facial features extracted from photo \\ \cline{2-4} 
    & Processing Approach    & Data processing implementation details & Facial recognition algorithm deployed on server \\ \hline
\multirow{3}{*}{\vspace{-.3cm}Data Access}      
    & Accessed Raw Data      & Collected raw data to be accessed & SSN, transaction history, photo \\ \cline{2-4} 
    & Accessed Output        & Processed data to be accessed & Risk score, facial features \\ \cline{2-4} 
    & Access Approach        & Data access details & Meeting host views attendees' attention scores in system-generated report \\ \hline
\multirow{3}{*}{\vspace{-.5cm}Choice \& Notice} 
    & Choice Options     & Options for the choice maker & User selects ``accept'' to allow data use for marketing or ``reject'' to deny it \\ \cline{2-4} 
    & Choice Impacts       & Effects of the user’s choice & ``Accept'' enables data scientist to use data in marketing analysis \\ \cline{2-4} 
    & Choice Notification & System notification upon/after choice & Pop-up message such as ``Your permission has been saved'' \\ \hline
\end{tabular}}
\label{tab:design_dec}
\end{table*}

\section{Patterns Occurred in Creator-Interpreter Communication}


\begin{table*}[ht]
\caption{We summarized six patterns in the creator-interpreter communication regarding each design decision. We present sample cases to illustrate each pattern. For instance, if the creator annotated the content of a design decision (e.g., \textit{Collected Personal Data}) as ``SSN'', while the interpreter didn't annotate, we would code the pattern as (E) Only the creator responded.}
\centering
\resizebox{.75\linewidth}{!}{%
\begin{tabular}{|c|l|c|c|}
\hline
\multicolumn{1}{|c|}{\textbf{\#}} &
  \multicolumn{1}{c|}{\textbf{Pattern}} &
  \multicolumn{1}{c|}{\textbf{Creator's Response}} &
  \multicolumn{1}{c|}{\textbf{Interpreter's Response}} \\ \hline\hline
A  & Full alignment &  SSN, transaction history  &  SSN, transaction history     \\ \hline
B & Partial alignment & SSN, transaction history &  SSN, date of birth\\ \hline
C & No alignment & SSN & Transaction history  \\ \hline
D & Only the interpreter responded & -- &  SSN   \\ \hline
E  & Only the creator responded & SSN & --  \\ \hline
F  & Neither responded & -- & --    \\ \hline
\end{tabular}}\label{tab:six-pattern}
\end{table*}

\newpage

\section{Definition of Interpretation Recall and Precision in Section~\ref{sec:interp-method-analysis}}\label{apdx:def-recall-precison}

\begin{table*}[htb]

\caption{We define precision, recall, and F1 score based on the six patterns occurred in creator-interpreter communication. Specifically, the weight of partial alignment (\(\epsilon\)) we used in Table~\ref{tab:ele_cover_rate} is 0.5. We also validated our results with \(\epsilon\)~=~0 in Table~\ref{tab:ele_cover_rate_0}.}
\centering
\resizebox{.95\linewidth}{!}{%
\begin{tabular}{|m{.128\linewidth}|m{.41\linewidth}|m{.18\linewidth}|}
\hline
\multicolumn{1}{|c|}{\textbf{Term}} & \multicolumn{1}{c|}{\textbf{Explanation}} & \multicolumn{1}{c|}{\textbf{Formula}} \\
\hline\hline
True Positive (TP) &
Interpreter's response fully or partially aligned with the creator's response (partial alignment weighted by \(\epsilon\)). &
\(\#A + \epsilon \cdot \#B\) \\
\hline
False Positive (FP) &
Interpreter responded when the creator did not, including the unaligned  (\(1 - \epsilon\)) part of \(B\). &
\(\#D + (1 - \epsilon) \cdot \#B\) \\
\hline
False Negative (FN) &
The creator responded, but the interpreter either did not respond or misinterpreted the response. &
\(\#C + \#E\) \\
\hline
True Negative (TN) &
Neither the creator nor the interpreter responded. &
\(\#F\) \\
\hline
\textbf{Precision} &
The proportion of correctly aligned responses out of all responses made by the interpreter. &
\(\frac{\#A + \epsilon \cdot \#B}{\#A + \epsilon \cdot \#B + \#D + (1 - \epsilon) \cdot \#B}\) \\
\hline
\textbf{Recall} &
The proportion of correctly aligned responses out of all responses made by the creator. &
\(\frac{\#A + \epsilon \cdot \#B}{\#A + \epsilon \cdot \#B + \#C + \#E}\) \\
\hline
\textbf{F1 Score} &
The harmonic mean of precision and recall, balancing the two. &\vfill\vspace{.14cm}
\(2 \cdot \frac{\textit{Precision} \cdot \textit{Recall}}{\textit{Precision} + \textit{Recall}}\) \vfill\\
\hline
\end{tabular}
}
\label{apdx:confusion-mat}
\end{table*}


\begin{table*}[h]
\centering
\caption{In Table~\ref{tab:ele_cover_rate}, we used a partial alignment weight of $\epsilon = 0.5$ to calculate interpretation precision, recall, and F1. To prevent potential bias, we repeated the analysis with $\epsilon = 0$ (i.e., treating ``partial alignment'' as ``no alignment''). This table shows that the differences between the two conditions remained consistent with our previous results, indicating our findings in Section~\ref{sec:interp-method-analysis} are valid. For each metric, we \change{highlight the higher value and~}indicate the statistical significance between the two conditions using the Wilcoxon Signed-Rank tests ($^{*}p~\textless~0.05$, $^{**}p~\textless~0.01$, $^{***}p~\textless~0.001$)\change{, all showing large effect sizes ($r~\textgreater~0.5$)}.}
\resizebox{.75\linewidth}{!}{%
\begin{tabular}{|p{3cm}|>{\centering\arraybackslash}p{1.1cm}>{\centering\arraybackslash}p{.8cm}>{\centering\arraybackslash}p{.8cm}|>{\centering\arraybackslash}p{1.1cm}>{\centering\arraybackslash}p{.8cm}>{\centering\arraybackslash}p{.8cm}|}
\hline
\multicolumn{1}{|c|}{\multirow{2}{*}{Design Decision}} & \multicolumn{3}{c|}{Vocabulary-Guided Sketches}                   & \multicolumn{3}{c|}{Heuristic-Guided Sketches}              \\
\multicolumn{1}{|c|}{}                         & Precision & Recall & F1 & Precision & Recall & F1 \\ \hline\hline
Collected Personal Data &
  81.0\% &
  89.5\% &
  85.0\% &
  \textbf{90.9\%} &
  \textbf{100.0\%} &
  \textbf{95.2\%} \\
Data Provider &
  59.1\% &
  61.9\% &
  60.5\% &
  \textbf{87.0\%} &
  \textbf{90.9\%} &
  \textbf{88.9\%} \\
Collection Purpose &
  22.2\% &
  23.5\% &
  22.9\% &
  \textbf{50.0\%} &
  \textbf{58.8\%} &
  \textbf{54.1\%} \\
Stored Data &
  18.8\% &
  17.6\% &
  18.2\% &
  \textbf{77.8\%} &
  \textbf{77.8\%} &
  \textbf{77.8\%} \\
Storage Approach &
  27.8\% &
  35.7\% &
  31.3\% &
  \textbf{68.4\%} &
  \textbf{76.5\%} &
  \textbf{72.2\%} \\
Post-Storage Action &
  6.7\% &
  14.3\% &
  9.1\% &
  \textbf{37.5\%} &
  \textbf{42.9\%} &
  \textbf{40.0\%} \\
Processing Input &
  23.5\% &
  22.2\% &
  22.9\% &
  \textbf{55.0\%} &
  \textbf{61.1\%} &
  \textbf{57.9\%} \\
Processing Output &
  42.1\% &
  44.4\% &
  43.2\% &
  \textbf{82.4\%} &
  \textbf{93.3\%} &
  \textbf{87.5\%} \\
Processing Approach &
  16.7\% &
  18.2\% &
  17.4\% &
  \textbf{53.3\%} &
  \textbf{57.1\%} &
  \textbf{55.2\%} \\
Accessed Raw Data &
  14.3\% &
  16.7\% &
  15.4\% &
  \textbf{54.5\%} &
  \textbf{50.0\%} &
  \textbf{52.2\%} \\
Accessed Output &
  \textbf{41.7\%} &
  35.7\% &
  38.5\% &
  38.9\% &
  \textbf{50.0\%} &
  \textbf{43.8\%} \\
Access Approach &
  27.3\% &
  23.1\% &
  25.0\% &
  \textbf{53.3\%} &
  \textbf{50.0\%} &
  \textbf{51.6\%} \\
Choice Options &
  36.4\% &
  36.4\% &
  36.4\% &
  \textbf{57.1\%} &
  \textbf{66.7\%} &
  \textbf{61.5\%} \\
Choice Impacts &
  0.0\% &
  0.0\% &
  0.0\% &
  \textbf{33.3\%} &
  \textbf{57.1\%} &
  \textbf{42.1\%} \\
Choice Notification &
  16.7\% &
  25.0\% &
  20.0\% &
  \textbf{66.7\%} &
  \textbf{66.7\%} &
  \textbf{66.7\%} \\ \hline\hline
  \multicolumn{1}{|c|}{\multirow{1}{*}{Mean}}    & 28.9\% & 30.9\% & 31.8\% & \textbf{60.4\%} & \textbf{66.6\%} & \textbf{63.1\%} \\ 
  \multicolumn{1}{|c|}{\multirow{1}{*}{$p$-value}}  & $^{***}$ & $^{***}$ & $^{***}$ & $^{***}$ & $^{***}$ & $^{***}$ \\ \hline

  \hline
\end{tabular}}
\label{tab:ele_cover_rate_0}
\end{table*}

\nobalance

\onecolumn

\newpage 

\section{Meta-Review}

The following meta-review was prepared by the program committee for the 2025
IEEE Symposium on Security and Privacy (S\&P) as part of the review process as
detailed in the call for papers.

\subsection{Summary}
The paper presents a user study of n=24 data science students (both undergraduate and graduate) who were asked to sketch privacy designs of a set of privacy scenarios after being given three heuristics to follow at the beginning of the study. The paper describes two pre-studies. The first pre-study was a need-finding study (n=12) that found that students lacked the vocabulary needed to sketch privacy designs effectively, the sketching may require multiple iterations, and the students had difficulty planning the sketch space in advance. The second pre-study is a teaching methods experiment that investigates 3 methods: object-oriented diagramming, vocabulary-based sketching, and heuristic-based sketching. The findings suggested that heuristic-based and vocabulary-based sketching would be the best for data science students, and those methods are examined in the main study. The main study (n=24) compared the two methods in a sketching task, followed by a post-task interview. The findings suggest that the simple heuristic method performed best in sketching efficiency and broader coverage of privacy-related design decisions and had a reduced learning curve. A heuristic-based approach combined with the multi-layer representation is recommended for privacy design sketches conducted by data science students.

\subsection{Scientific Contributions}
\begin{itemize}
\item Provides a Valuable Step Forward in an Established Field
\item Establishes a New Research Direction
\end{itemize}

\subsection{Reasons for Acceptance}
\begin{enumerate}
\item Provides a Valuable Step Forward in an Established Field. The paper contributes to the research on sketching privacy designs and offers insights into how we should teach data science students about data privacy design.
\item Establishes a New Research Direction. This paper presents the design and evaluation of a novel approach to prompting data science students to consider privacy in their design. The proposed heuristic-based prompting offers a novel direction of investigation for future work.
\end{enumerate}




\end{document}